\documentclass[times,10pt]{article}  
\usepackage{amssymb}     
\usepackage{amsfonts}     
\usepackage{latexsym}       
\usepackage{epsfig}   
\usepackage{diagrams}    
\usepackage{verbatim} 
\usepackage{times}      

\newarrow{Is}=====
\newcommand{\suck}{\vspace{-1mm}}

\newtheorem{Th}{Theorem}[section]

\newtheorem{definition}[Th]{Definition}

\newcommand{\bit}{\begin{itemize}} 
\newcommand{\eit}{\end{itemize}\par\noindent} 
\newcommand{\ben}{\begin{enumerate}} 
\newcommand{\een}{\end{enumerate}\par\noindent} 
\newcommand{\beq}{\begin{equation}} 
\newcommand{\eeq}{\end{equation}\par\noindent} 
\newcommand{\beqa}{\begin{eqnarray*}} 
\newcommand{\eeqa}{\end{eqnarray*}\par\noindent}  
\newcommand{\beqn}{\begin{eqnarray}}   
\newcommand{\eeqn}{\end{eqnarray}\par\noindent}

\def\PP{{\rm P}}  
\def\HH{{\cal H}}  
\def\CC{{\bf C}}  
\def\II{{\rm I}}  
\newcommand{\dd}{\llcorner}
\newcommand{\sdot}{\bullet}
\newcommand{\ddd}{\lrcorner}
\newcommand{\uu}{\ulcorner} 
\newcommand{\uuu}{\urcorner}

\title{\bf Quantum information-flow,\\ concretely, and axiomatically} 
\author{
Bob Coecke\footnote{Howard Barnum, Rick Blute, Sam Braunstein, 
Vincent Danos, Ross Duncan, Peter Hines, Martin Hyland,
Prakash Panangaden, Peter Selinger and Vlatko Vedral provided feedback. Samson Abramsky and
Mehrnoosh Sadrzadeh read this manuscript.}\\ 
{\small Oxford University Computing
Laboratory,}\\
{\small Wolfson Building, Parks  
Road, Oxford OX1 3QD, UK.}\\ 
\small\texttt{http://se10.comlab.ox.ac.uk:8080/BobCoecke/Home$\_$en.html}\\
{\small\texttt{bob.coecke@comlab.ox.ac.uk}}
}     

\date{} 
 
\begin{document}    
\maketitle 
 
\vspace{-5.4mm}\noindent 
\begin{abstract}
These lecture notes survey joint work with Samson Abramsky.  I will somewhat
informally discuss the main results of the papers 
\cite{Abr2,AbrCoe1,AbrCoe2,AbrCoe3,Coe1,Coe2}   
in a pedestrian not too technical way. These include: 
\bit
\item \em `The logic of entanglement' \em \cite{Coe2}, that is, the identification
and abstract axiomatization of the \em `quantum information-flow' \em which enables 
protocols such as quantum teleportation. To this means we
define
\em strongly compact closed categories \em which abstractly capture the behavioral
properties of quantum entanglement.
\item `\em Postulates for an abstract quantum formalism' \em \cite{AbrCoe2} in
which \em classical information-flow \em (e.g.~token exchange) is part of the
formalism. As an example, we provide a purely formal description of quantum
teleportation and \em prove correctness in abstract generality\em. In this
formalism
\em types reflect kinds\em, contra the essentially typeless von Neumann formalism
\cite{vN}.  Hence even concretely this formalism manifestly improves on the usual
one.
\item \em `Towards a high-level approach to quantum informatics' \em \cite{Abr2}.
Indeed, the above discussed work can be conceived as aiming to solve:

{\large\[ 
{{\bf ???}\over{\bf von\ Neumann\ quantum\ formalism}}
\ \simeq\  
{{\bf high\mbox{\bf-}level\ language}\over{\bf low\mbox{\bf-}level\ language}}\,. 
\]}

\eit
\end{abstract}

\section{What? When? Where? Why?} 

First of all, for us `quantum' stands for the concepts (both
operational and formal) which had to be added to classical physics in order to
understand observed phenomena such as the structure of the spectral lines in atomic
spectra, experiments exposing non-local correlations, seemingly $4\pi$ symmetries,
etc.  While the basic part of classical mechanics deals with the (essentially)
reversible unitary dynamics of physical systems, quantum required adding the notions
of measurement and (possibly non-local) correlations to the discussion. The
corresponding mathematical formalism was considered to have reached its maturity in
von Neumann's book
\cite{vN}. However!  

\paragraph{The quantum teleportation protocol.} The quantum
teleportation 
protocol
\cite{BBC} involves three qubits $a$, $b$ and $c$ and two spatial
regions
$A$ (for ``Alice'') and $B$ (for ``Bob''). 
Qubit $a$ is in a state
$|\phi\rangle$ and located in $A$.
Qubits $b$ and $c$ form an `EPR-pair', that is, their joint state is
$|00\rangle+|11\rangle$. We assume that these qubits are
initially in $B$ e.g.~Bob created them. 
After \emph{spatial relocation} so that $a$ and $b$ are located in $A$, while
$c$ is positioned in $B$, or in other words, ``Bob sends qubit $b$ to
$Alice$'', we can start the actual teleportation of qubit $a$.

\vspace{2.5mm}\noindent{
\hspace{-15pt}\begin{minipage}[b]{1\linewidth}
\centering{\epsfig{figure=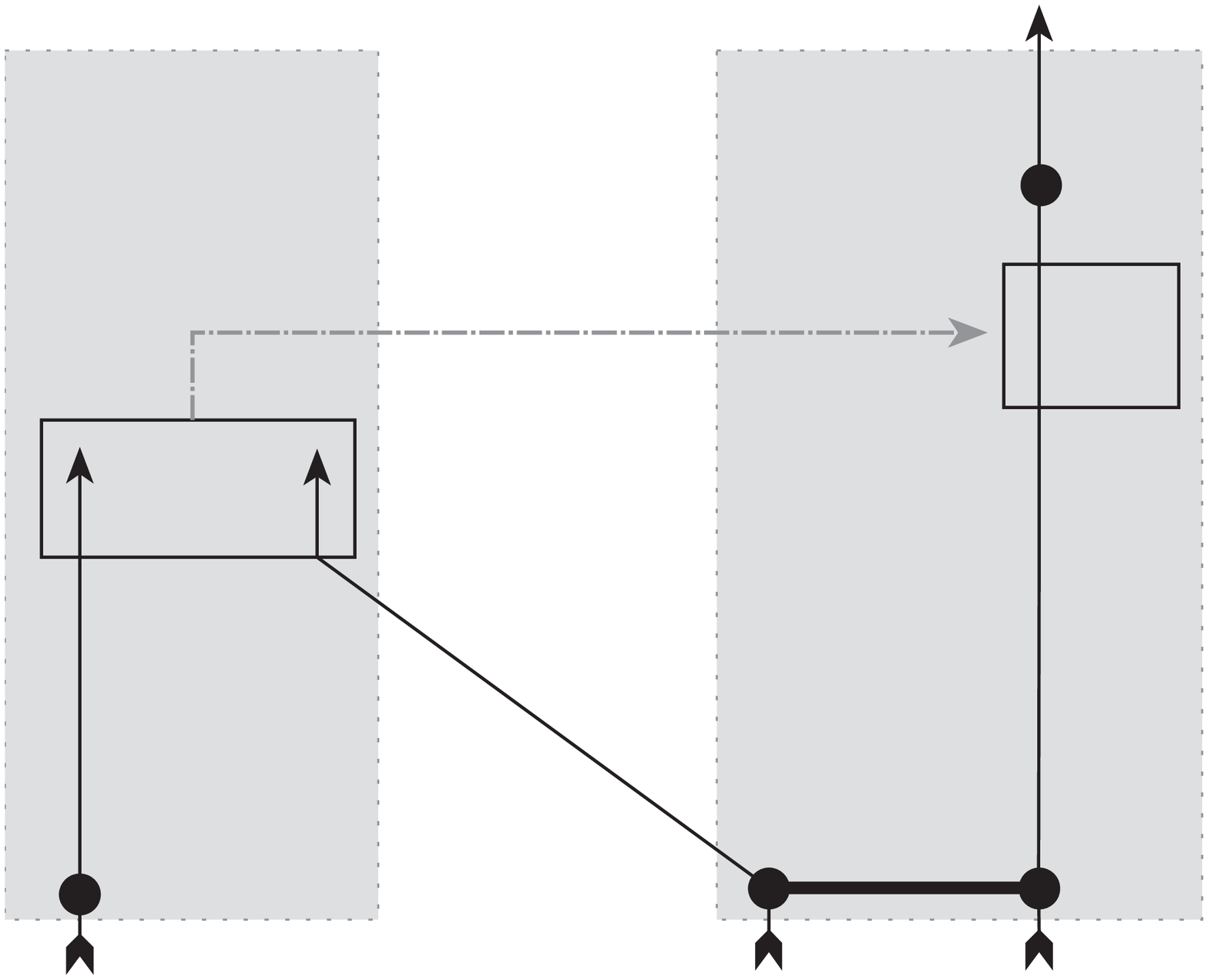,width=245pt}}

\hspace{3mm}\begin{picture}(245,0)
\put(155.5,37){\large$|00\rangle\!+\!|
\hspace{-0.5pt}1\hspace{-0.5pt}1\hspace{-0.5pt}\rangle$}
\put(20,108){\Large$M_{\!Bell}$}
\put(214,139){\Large$U_{\!x}$}
\put(91.5,149){\large${x\in\mathbb{B}^2}$}
\put(209.1,182.2){\large$|\phi\rangle$}
\put(14,37){\large$|\phi\rangle$}
\put(255,78){\vector(0,1){60}}
\put(257,104){${\rm time}$}
\put(-1,189){\Large$A$}
\put(144,189){\Large$B$}
\put(7.4,1){\Large$a$}
\put(149,0){\Large$b$}
\put(203.4,1){\Large$c$}
\end{picture}
\end{minipage}}

\vspace{1.5mm}\noindent
Alice performs a \em Bell-base measurement
\em $M_{Bell}$ on $a$ and $b$ at $A$, that is, a measurement such that each 
projector in the spectral decomposition of the corresponding
self-adjoint 
operator  projects on one of the
one-dimensional subspaces spanned by a vector in the \em
Bell basis\em: 
\[
b_1:={|00\rangle\!+\!|11\rangle\over \sqrt{2}}\quad
b_2:={|01\rangle\!+\!|10\rangle\over \sqrt{2}}\quad
b_3:={|00\rangle\!-\!|11\rangle\over \sqrt{2}}\quad
b_4:={|01\rangle\!-\!|10\rangle\over \sqrt{2}}\,.\!\!\!
\]
We will omit scalar multiples from now on.
This measurement may be of the `distructive' kind. Alice observes
the outcome of the measurement and ``sends these two classical bits
($x\in\mathbb{B}^2$) to
Bob''. Depending on which classical bits he receives Bob then
performs one of the unitary
transformations
\[
\beta_1:=\left(\begin{array}{rr} 
1&0\\
0&1
\end{array}\right)
\quad
\beta_2:=\left(\begin{array}{rr}
0&1\\
1&0
\end{array}\right)
\quad
\beta_3:=\left(\begin{array}{rr}
1&0\\ 
0&\!\!\!\!\!-\!1
\end{array}\right)
\quad
\beta_4:=\left(\begin{array}{rr}
0&\!\!\!\!\!-\!1\\
1&0
\end{array}\right) 
\]
on $c$ --- $\beta_1,\beta_2,\beta_3$
are all self-inverse while
$\beta_4^{-1}=-\beta_4$. 
The final state of $c$ proves to be $|\phi\rangle$ as well. 

\vspace{-1mm}
\paragraph{Where does ``it'' flow?} 
Consider this quantum teleportation protocol. In this process
continuous data is transmitted from Alice to Bob 
while only using a two-bit classical channel.
So where does the `additional information' flow?  The quantum formalism does not tell
us in an explicit manner.  Clearly it has something to do with the nature of
quantum compoundness, but, what exactly? Note that this reasonably simple protocol
was only discovered some 60 years after von Neumann's formalism.  \em Wouldn't it be
nice to have a formalism in which inventing quantum teleportation would be an
undergraduate exercise? \em

\vspace{-1mm}
\paragraph{Where are the types?} While in the lab measurements are applied to
physical systems, application of the corresponding self-adjoint operator
$M:{\cal H}\to{\cal H}$ to the vector $\psi\in{\cal H}$ which represents the system's
state, hence  yielding $M(\psi)$, does not reflect how the state changes
during the act of measurement!  The actual change is $\psi\mapsto \PP_i(\psi)$ for 
spectral decomposition 
$M=\sum_i a_i\cdot\PP_i$, where
$a_i$ is the outcome of the measurement.  In addition to this change of state a measurement involves provision of data
to `the observer' cf.~teleportation where this data determines the choice of the unitary correction. This 
contradicts what the corresponding types seem to indicate. The same argument goes for
the composite of two self-adjoint operators which in general is not self-adjoint while
measurements can be performed sequentially in the lab.  \em Wouldn't it be nice 
if types reflect kinds? \em 

\vspace{-1mm} 
\paragraph{Much worse even, where is the classical information and its flow?}
Indeed, the problem regarding types is directly connected to the fact that in von
Neumann's formalism there is no place for storage, manipulation and exchange of the
classical data obtained from measurements.  \em We want a
quantum formalism which allows to encode classical information and its flow, and hence also one
which has enough types to reflect this!
\em

\vspace{-1mm}
\paragraph{What is the true essence of quantum?} John von Neumann himself was the
first to look for this, teaming up with the  `king of lattices' Garrett Birkhoff
\cite{BvN}.  It is fair
to say that as an attempt to understand `the whole of quantum
mechanics' this particular `quantum logic' program has
failed.  While it provided a much better understanding of
quantum superposition and the superselection rules (for a
survey try to get hold of Piron's
\cite{Piron} and Varadarajan's \cite{var} books), it failed
at teaching us anything about quantum entanglement, and
definitely didn't teach us anything on how quantum and
classical information interact.  So lattices don't seem to
be capable of doing the job.  \em Which mathematical setting
provides an abstract  quantum formalism, and its
corresponding logic? \em   
 
\section{The logic of entanglement} 

\paragraph{A mathematics exercise.} The `Where does ``it'' flow?' question was addressed and solved in \cite{Coe1,Coe2}.
But the result challenges quantum mechanics' faithfulness to vector spaces!
 We start by playing a quiz testing the reader's
knowledge on the Hilbert space tensor product. Consider the situation depicted below
where all boxes represent bipartite projectors on one-dimensional subspaces of Hilbert spaces
${\cal H}_i\otimes{\cal H}_j$, that is, linear maps
\[
\PP_\Xi:{\cal H}_i\otimes{\cal H}_j\to{\cal H}_i\otimes{\cal
H}_j::\Phi\mapsto\langle\Psi_\Xi\mid\Phi\rangle\cdot\Psi_\Xi
\]
with $\Psi_\Xi\in{\cal H}_i\otimes{\cal H}_j$ and $|\Psi_\Xi|=1$ so
$\PP_\Xi(\Psi_\Xi)=\Psi_\Xi$, $\phi_{in}\in{\cal H}_1$, $\phi_{out}\in{\cal
H}_5$, $\Phi_{{in}}\in{\cal H}_2\otimes{\cal H}_3\otimes{\cal H}_4\otimes{\cal H}_5$ and
hence
$\Psi_{\bf in},\Psi_{\bf out}\in{\cal H}_1\otimes{\cal H}_2\otimes{\cal H}_3\otimes{\cal
H}_4\otimes{\cal H}_5$,   

\vspace{9mm}\noindent\begin{minipage}[b]{1\linewidth}  
\centering{\epsfig{figure=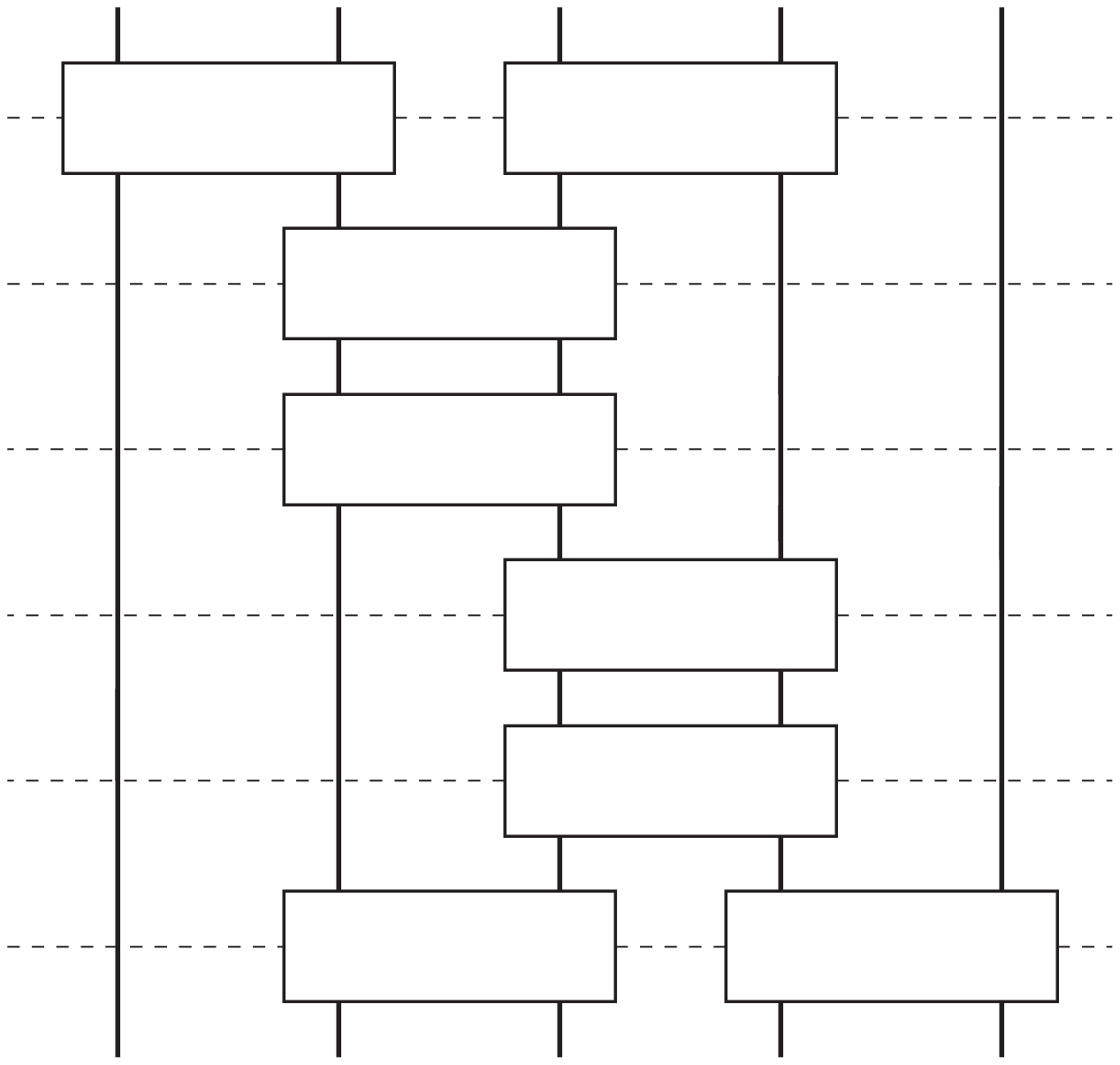,width=133.3pt}}    

\vspace{3.3mm}
\begin{picture}(133.3,0)   
\put(35,28){$\underbrace{\mbox{\hspace{3.3cm}}}$} 
\put(-28,16){\footnotesize$\Psi_{\bf in}:=$}
\put(12,16){\footnotesize$\phi_{in}$}
\put(79,15){\scriptsize$\Phi_{{in}}$}
\put(-28,164,5){\footnotesize$\Psi_{\bf out}:=$}
\put(115.5,164.5){\footnotesize$\phi_{out}$?}  
\put(24,164.5){\footnotesize$\Psi_{\rm V\!I\!I}$} 
\put(77,164.5){\footnotesize$\Psi_{\rm V\!I\!I\!I}$} 
\put(6,155.5){$\overbrace{\mbox{\hspace{1.6cm}}}$} 
\put(59,155.5){$\overbrace{\mbox{\hspace{1.6cm}}}$} 
\put(77,139.5){\footnotesize$\PP_{\rm V\!I\!I\!I}$}
\put(24,139.5){\footnotesize$\PP_{\rm V\!I\!I}$}
\put(50.5,119.5){\footnotesize$\PP_{\rm V\!I}$}
\put(51,99.5){\footnotesize$\PP_{\rm V}$}
\put(77,79.5){\footnotesize$\PP_{\rm I\!V}$}
\put(77,59.5){\footnotesize$\PP_{\rm I\!I\!I}$}
\put(103.5,39.5){\footnotesize$\PP_{\rm I\!I}$} 
\put(51,39.5){\footnotesize$\PP_{\rm I}$}
\end{picture} 
\vspace{0mm}
\end{minipage}   

\vspace{-0.5mm}\noindent
{\LARGE\begin{center} 
\fbox{What is $\phi_{out}$?} 
\end{center}}

\vspace{1mm}\noindent
(up to a scalar multiple is ok!)~~In algebraic terms this
means solving
\[
k\cdot \zeta\left(\phi_{in}\,\otimes\Phi_{{in}}\right)=\Psi_{\rm V\!I\!I}\otimes\Psi_{\rm
V\!I\!I\!I}\otimes
\phi_{out}
\]
in the unknown $\phi_{out}$ for $k\in\mathbb{C}$ and
\beqa
\zeta\!\!&:=&\!\!
(\PP_{\rm V\!I\!I}\otimes\PP_{\rm V\!I\!I\!I}\otimes 1_5)\circ
(1_1\otimes\PP_{\rm V\!I}\otimes 1_{4,5})\circ
(1_1\otimes\PP_{\rm V}\otimes 1_{4,5})\circ\\
&&\!\!\quad\quad\ 
(1_{1,2}\otimes\PP_{\rm I\!V}\otimes 1_5)\circ
(1_{1,2}\otimes\PP_{\rm I\!I\!I}\otimes 1_5)\circ
(1_1\otimes \PP_{\rm I}\otimes\PP_{\rm I\!I})
\eeqa 
where $1_i$ is the identity on ${\cal H}_i$ and $1_{ij}$ is the identity on ${\cal H}_i\otimes{\cal H}_j$.

At first sight this seems a randomly chosen nasty problem without conceptual significance. But it is not!
Observe that bipartite vectors $\Psi\in{\cal H}_1\otimes{\cal
H}_2$ are in bijective correspondence with linear maps $f:{\cal H}_1\!\to\!{\cal
H}_2$ through matrix representation in bases $\{e_{i}^{(1)}\}_i$ and
$\{e_{j}^{(2)}\}_j$ of ${\cal H}_1$ and
${\cal H}_2$,
\[
\Psi=\!\sum_{ij} m_{ij}\cdot e_{i}^{(1)}\!\otimes e_{j}^{(2)}\ 
\stackrel{\simeq}{\longleftrightarrow}\
\left(\begin{array}{ccc}\vspace{-1mm}
\!\!m_{11}&\!\!\cdots&\!\!m_{1n}\!\!\\ \vspace{-1mm}
\!\!\vdots&\!\!\ddots&\!\!\vdots\!\!\\ 
\!\!m_{k1}&\!\!\cdots&\!\!m_{kn}\!\!
\end{array}\right)\
\stackrel{\simeq}{\longleftrightarrow}\
f::e_{i}^{(1)}\!\!\mapsto\!\sum_{j} m_{ij}\cdot e_{j}^{(2)}\!,
\]
or in bra-ket/qu-nit notation,
\[
\sum_{ij} m_{ij}\mid i\,j\,\rangle= 
\sum_{ij} m_{ij}\mid i\,\rangle\otimes\!\!\mid j\,\rangle\ 
\stackrel{\simeq}{\longleftrightarrow}\
\sum_{ij} m_{ij}\,\langle\, i\mid\!-\rangle\cdot\!\!\mid j\,\rangle\,.  
\]
This correspondence lifts to an isomorphism of vector spaces. 
As an example, the
(non-normalized) EPR-state corresponds to the identity
\[
\mid 00\,\rangle+\mid 11\,\rangle\
\stackrel{\simeq}{\longleftrightarrow}\
\left(\begin{array}{cc}
1&0\\
0&1
\end{array}\right)\
\stackrel{\simeq}{\longleftrightarrow}\
\mathbf{1}=\langle\,0\!\mid\!-\rangle\cdot\!\!\mid 0\,\rangle +
\langle\,1\!\mid\!-\rangle\cdot\!\!\mid 1\,\rangle\,.
\]
In fact, the correspondence between ${\cal H}_1\otimes{\cal H}_2$ and anti-linear maps 
is a more natural one, since it is independent on the choice of a base for ${\cal
H}_1$,
\[
\sum_{ij} m_{ij}\mid i\,j\,\rangle= 
\sum_{ij} m_{ij}\,\mid i\,\rangle\otimes\!\!\mid j\,\rangle\ 
\stackrel{\simeq}{\longleftrightarrow}\
\sum_{ij} m_{ij}\,\langle-\!\mid i\,\rangle\cdot\!\!\mid j\,\rangle\,,
\]
or equivalently, the correspondence between ${\cal H}_1^*\otimes{\cal H}_2$ and
linear maps, where ${\cal H}_1^*$ is the vector space of linear functionals $\varphi:{\cal
H}_1\to\mathbb{C}$ which arises by setting $\varphi:=\langle \psi\mid-\rangle$ for
each $\psi\in{\cal H}_1$.
We will ignore this for now (see \cite{Coe1} for a detailed
discussion) and come back to this issue later.  

Since we can now `represent' vectors $\Psi_\Xi\in {\cal H}_i\otimes{\cal H}_j$ by linear
functions
of type ${\cal H}_i\to{\cal H}_j$, and hence also
the projectors $\PP_\Xi$ which appear in the above picture, we can redraw that picture as 

\vspace{9.5mm}\noindent\begin{minipage}[b]{1\linewidth}  
\centering{\epsfig{figure=Q1.eps,width=133.3pt}}    

\vspace{3.3mm}
\begin{picture}(133.3,0)   
\put(35,28){$\underbrace{\mbox{\hspace{3.3cm}}}$} 
\put(-28,16){\footnotesize$\Psi_{\bf in}:=$}
\put(12,16){\footnotesize$\phi_{in}$}
\put(79,15){\scriptsize$\Phi_{{in}}$}
\put(-28,164,5){\footnotesize$\Psi_{\bf out}:=$}
\put(115.5,164.5){\footnotesize$\phi_{out}$?}  
\put(24,164.5){\footnotesize$\Psi_{\rm V\!I\!I}$} 
\put(77,164.5){\footnotesize$\Psi_{\rm V\!I\!I\!I}$} 
\put(6,155.5){$\overbrace{\mbox{\hspace{1.6cm}}}$} 
\put(59,155.5){$\overbrace{\mbox{\hspace{1.6cm}}}$} 
\put(12,141.5){$-\!\!\!\!-{\scriptstyle f_1}\!\to$}
\put(65,141.5){$-\!\!\!\!-{\scriptstyle f_3}\!\to$}
\put(38.5,118.5){$-\!\!\!\!-{\scriptstyle f_2}\!\to$}
\put(65,78.5){$\leftarrow\!{\scriptstyle f_4}-\!\!\!\!-$}
\put(39,101.5){$\leftarrow\!{\scriptstyle f_5}-\!\!\!\!-$}
\put(39,38.5){$-\!\!\!\!-{\scriptstyle f_6}\!\to$}
\put(65,61.5){$-\!\!\!\!-{\scriptstyle f_7}\!\to$}
\put(91.5,38.5){$-\!\!\!\!-{\scriptstyle f_8}\!\to$} 
\end{picture} 
\vspace{0mm}
\end{minipage}   

\vspace{-2mm}\noindent
where now $\Psi_{\rm V\!I\!I} \stackrel{\simeq}{\longleftrightarrow}f_1$ and $\Psi_{\rm
V\!I\!I\!I}
\stackrel{\simeq}{\longleftrightarrow}f_3$, and the arrows $-\!\!\!\!-{\scriptstyle 
f_i}\!\to$ specify the domain and the codomain of the functions $f_i$, and, I should
mention that the new (seemingly somewhat random) numerical labels of the functions and the
direction of the arrows are well-chosen (since, of course, I know the answer to the quiz
question). We claim that, provided
$k\not=0$ (see \cite{Coe1}),
\[
\phi_{out}=(f_8\circ f_7\circ f_6\circ f_5\circ f_4\circ 
f_3\circ f_2\circ f_1)(\phi_{in}) 
\]
(up to a scalar multiple),
and we also claim that this is due to the fact that
we can draw a `line' of which the allowed passages through a projector are restricted to

\vspace{1mm}\noindent
\begin{minipage}[b]{1\linewidth}   
\centering{\epsfig{figure=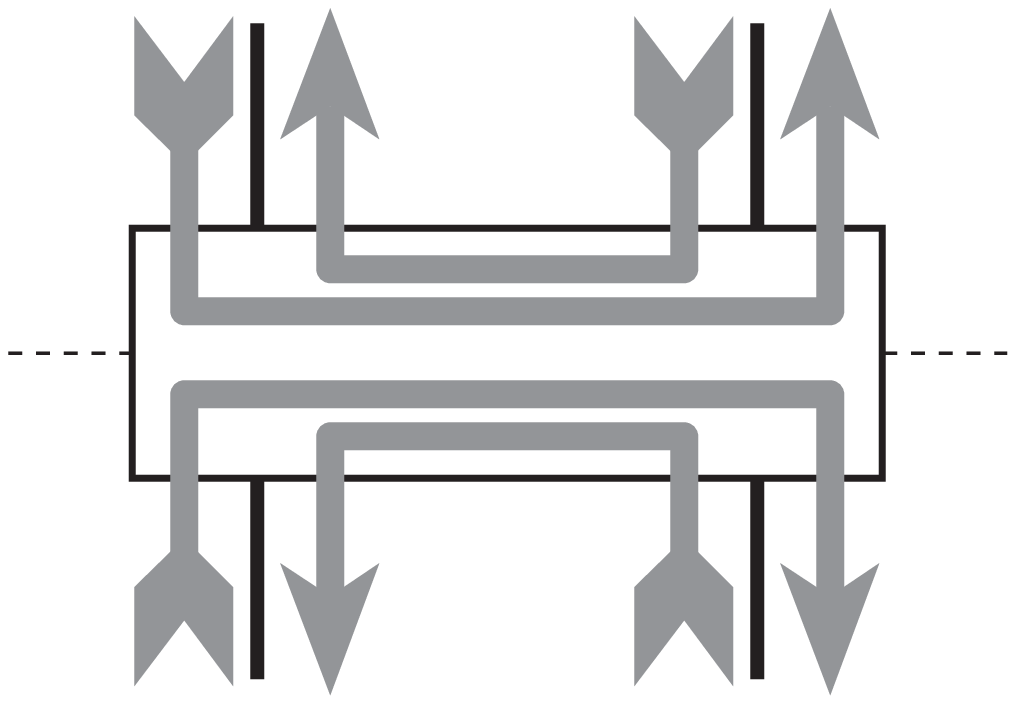,width=60pt}}   
\end{minipage}

\vspace{0.5mm}\noindent
that is, if the line enters at an input (resp.~output) of a bipartite box then it has
to leave by the other input (resp.~output) of that box (note the deterministic nature of
the path).  In other words

\medskip\noindent
\centerline{\bf\fbox{Permitted are:}}
\begin{center}
\epsfig{figure=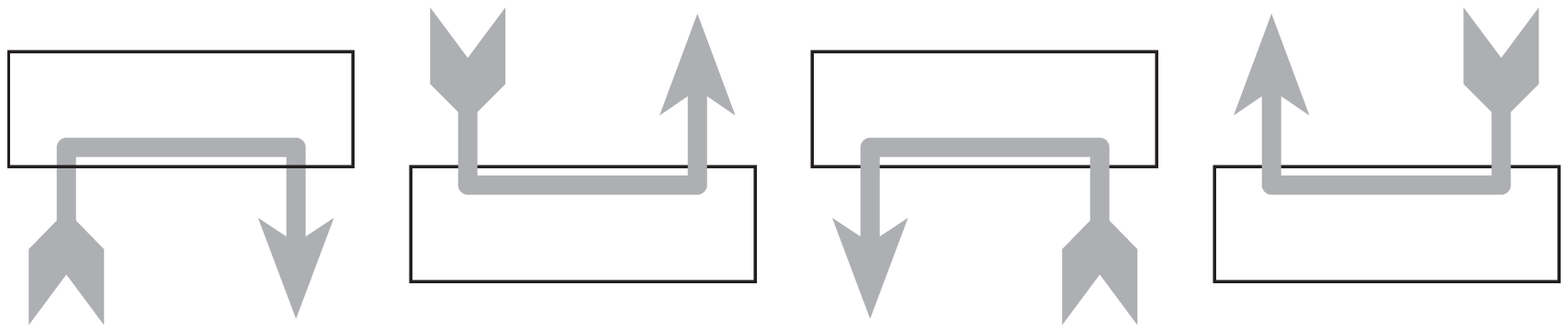,width=210pt}
\end{center} 
\centerline{\bf\fbox{Forbidden are:}}
\begin{center}
\epsfig{figure=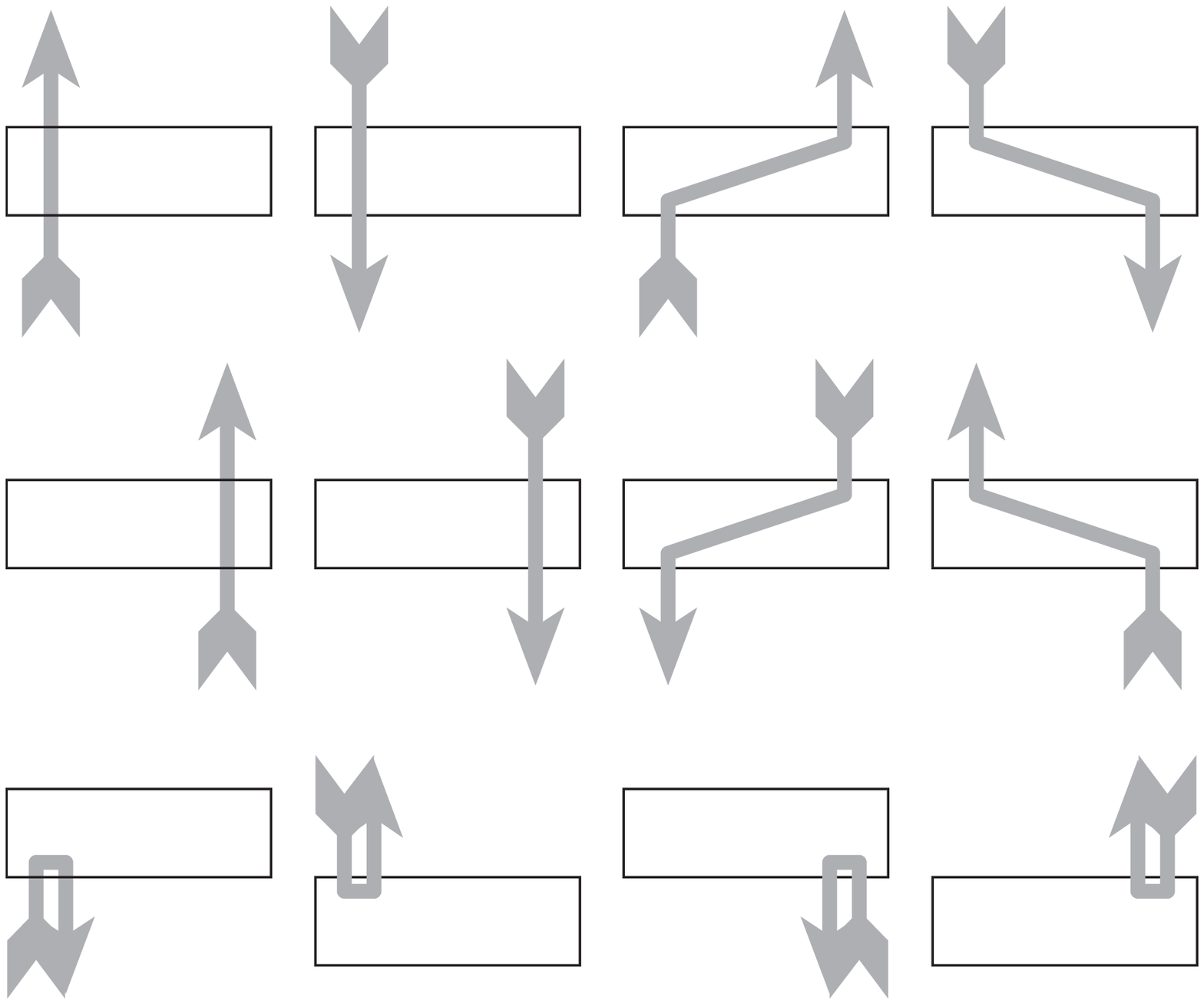,width=210pt}    
\end{center} 

\noindent
what results in:

\vspace{1.7mm}\noindent\begin{minipage}[b]{1\linewidth}    
\centering{\epsfig{figure=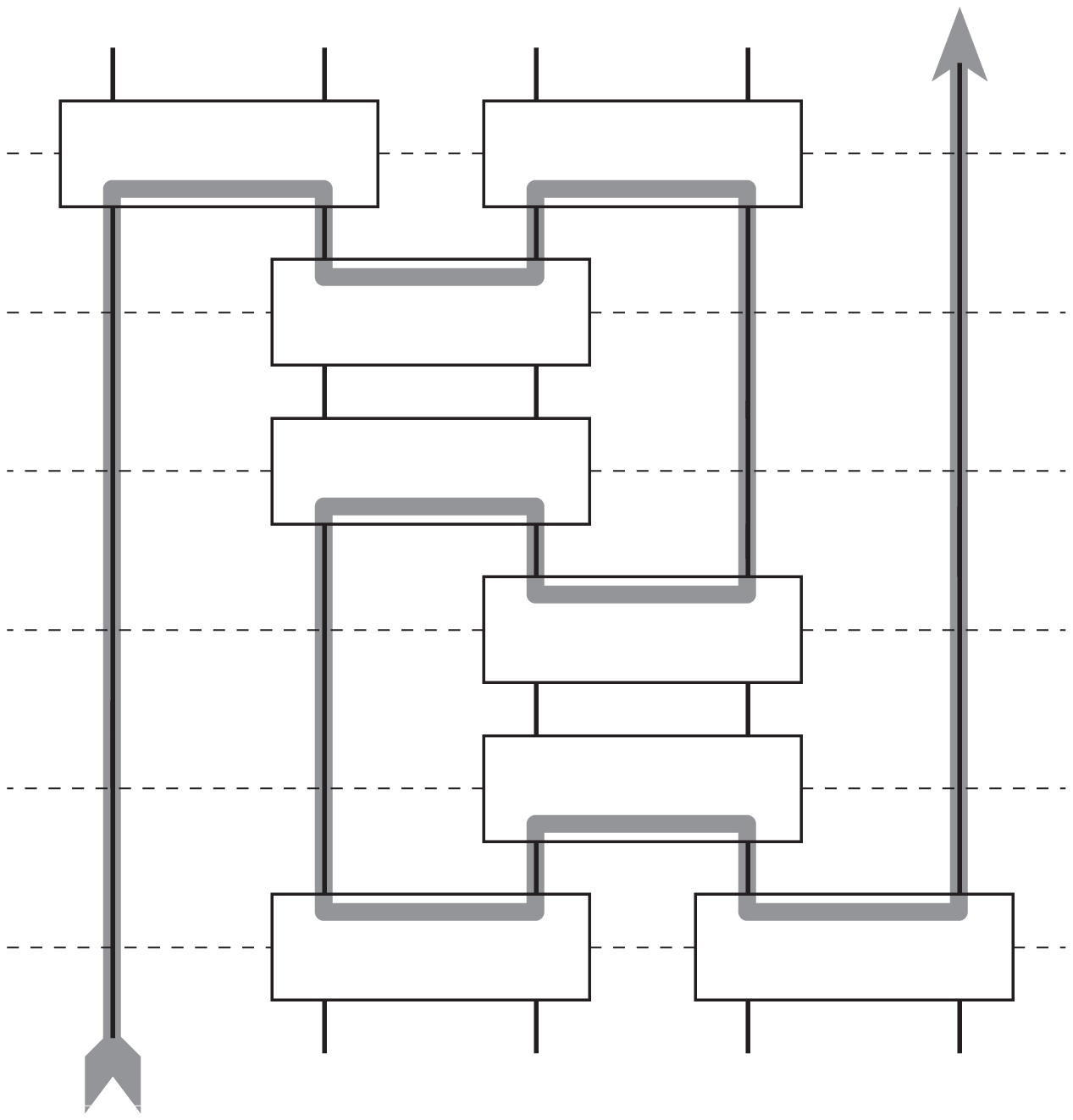,width=133.3pt}}    

\vspace{3.3mm}
\begin{picture}(133.3,0)   
\put(12,16){\footnotesize$\phi_{in}$}
\put(115.5,164.5){\footnotesize$\phi_{out}=(f_8\circ \ldots\circ f_1)(\phi_{in})$}  
\put(12,141.5){$-\!\!\!\!-{\scriptstyle f_1}\!\to$}
\put(65,141.5){$-\!\!\!\!-{\scriptstyle f_3}\!\to$}
\put(38.5,118.5){$-\!\!\!\!-{\scriptstyle f_2}\!\to$}
\put(65,78.5){$\leftarrow\!{\scriptstyle f_4}-\!\!\!\!-$}
\put(39,101.5){$\leftarrow\!{\scriptstyle f_5}-\!\!\!\!-$} 
\put(39,38.5){$-\!\!\!\!-{\scriptstyle f_6}\!\to$}
\put(65,61.5){$-\!\!\!\!-{\scriptstyle f_7}\!\to$}
\put(91.5,38.5){$-\!\!\!\!-{\scriptstyle f_8}\!\to$} 
\end{picture} 
\vspace{0mm}
\end{minipage}   

\vspace{-2.3mm}\noindent
When we follow this line, we first pass through the box labeled $f_1$, then the one
labeled $f_2$ and so on until
$f_8$.  Hence it seems {\bf ``as if''}
\em the information flows from $\phi_{in}$ to
$\phi_{out}$ following that line and that the functions $f_i$ labeling the boxes act on this
information\em. 
Also,  $\phi_{out}=(f_8\circ \ldots\circ f_1)(\phi_{in})$ does not depend on the input of
the projectors at
${\cal H}_2\otimes{\cal H}_3\otimes{\cal H}_4\otimes{\cal H}_5$ and, more importantly, the order
in which we apply the projectors does not reflect the order in which $f_1,
\ldots, f_8$ are applied to
$\phi_{in}$ in the expression $(f_8\circ \ldots\circ f_1)(\phi_{in})$.  Doesn't this have a
somewhat `acausal' flavor to it?

\vspace{-1.5mm}
\paragraph{The logic of quantum entanglement.} We claim that the above purely mathematical
observation exposes a \em quantum information-flow\em. It suffices to conceive the
projectors $\PP_\Xi$ as appearing in the spectral decompositions of self-adjoint operators
$M_\Xi:=\sum_i a_{\Xi,i}\cdot\PP_{\Xi,i}$ representing quantum measurements, that is, for
some $i$ we have 
$\PP_\Xi=\PP_{\Xi,i}$ (hence the outcome of the measurement represented by
$M_\Xi$ is
$a_{\Xi,i}$). As an example, consider

\vspace{4.5mm}\noindent{\footnotesize
\begin{minipage}[b]{1\linewidth}  
\centering{\epsfig{figure=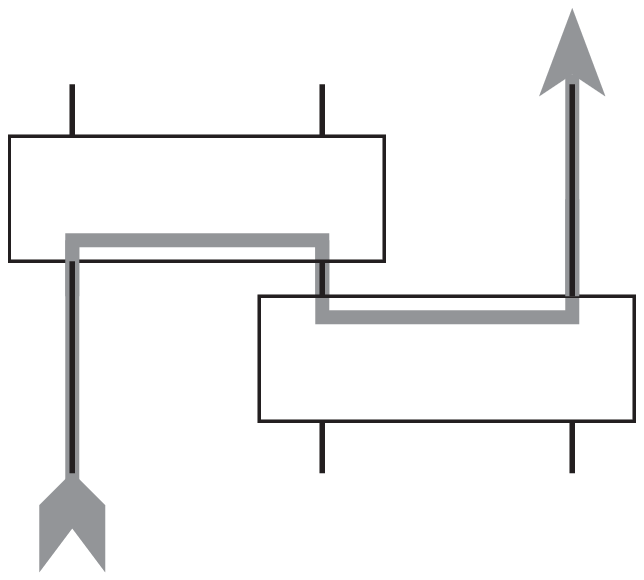,width=60pt}}   

\begin{picture}(160,0)   
\put(90,27){${\rm 1}$} 
\put(67.5,44){$\!{\rm 1}$}  
\put(52,4){\small$\phi_{in}$} 
\put(100,67){\small$\phi_{out}=(1\circ 1)(\phi_{in})=\phi_{in}$}  
\end{picture}   
\end{minipage}}

\vspace{-1mm}\noindent
where, since all labeling functions are identities, both projectors project on the
EPR-state. Since the first projector corresponds to `preparing an EPR-state', this picture
seems to provide us with a teleportation protocol, 

\vspace{2.5mm}\noindent{
\begin{minipage}[b]{1\linewidth}
\centering{\epsfig{figure=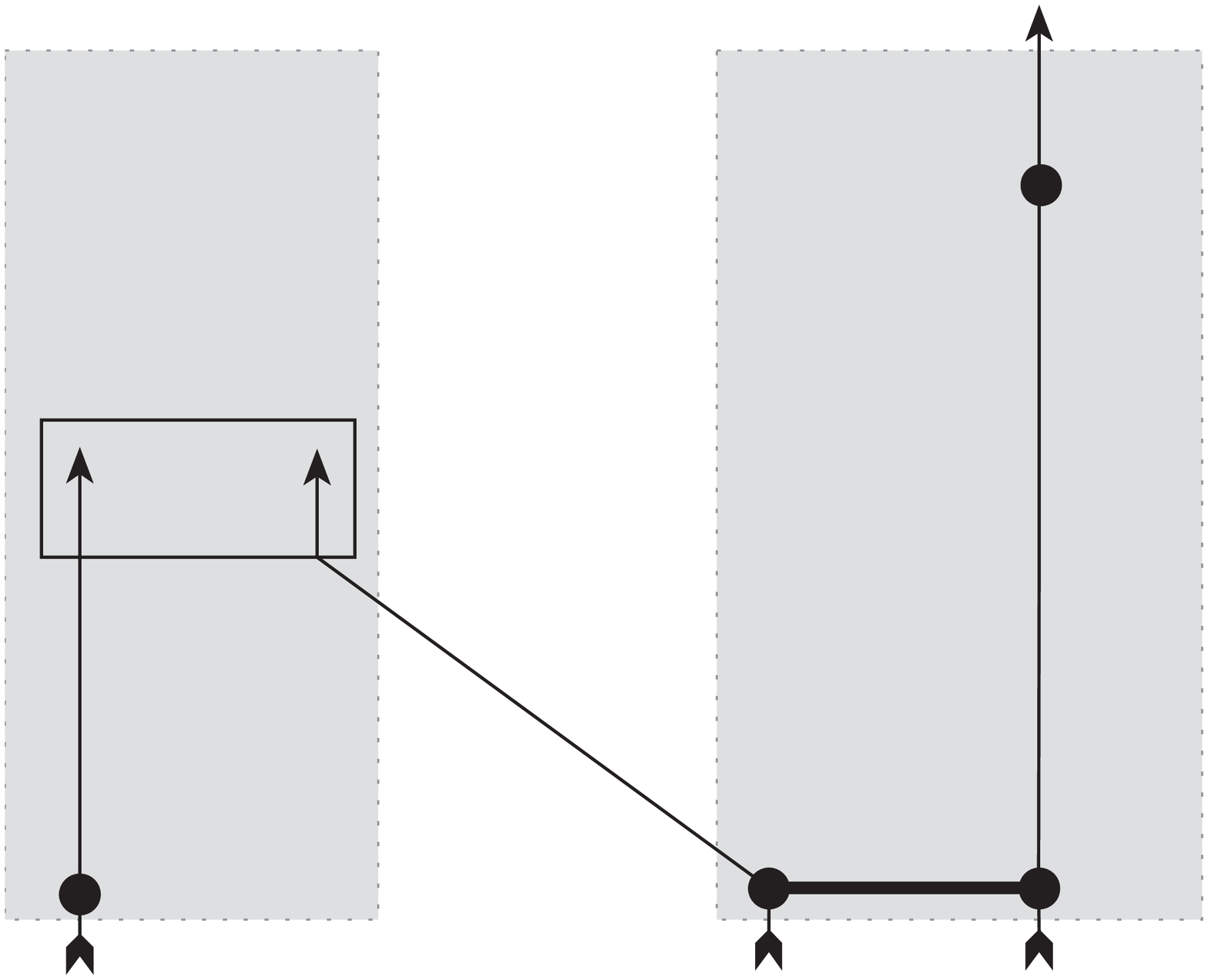,width=120pt}}

\hspace{3mm}\begin{picture}(120,0)
\put(79,25){\footnotesize EPR}
\put(7,58){\footnotesize${\rm
P}{\scriptstyle{}_{\!E\!P\!R}}$} 
\put(103,90){\footnotesize$|\phi\rangle$} 
\put(7,20){\footnotesize$|\phi\rangle$}
\end{picture}
\end{minipage}}

\vspace{1.5mm}\noindent
However, physically we cannot implement $\PP_{\!E\!P\!R}$ on its
own `with certainty'.   

But $\PP_{\!E\!P\!R}$ is part of Bell-base
measurement together with three other projectors. We denote the corresponding
labeling functions by
$\beta_2,\beta_3,\beta_4$. The grey boxes below denote unitary 
transformations. We have

\vspace{6mm}\noindent
\begin{minipage}[b]{1\linewidth}  
\centering{\epsfig{figure=
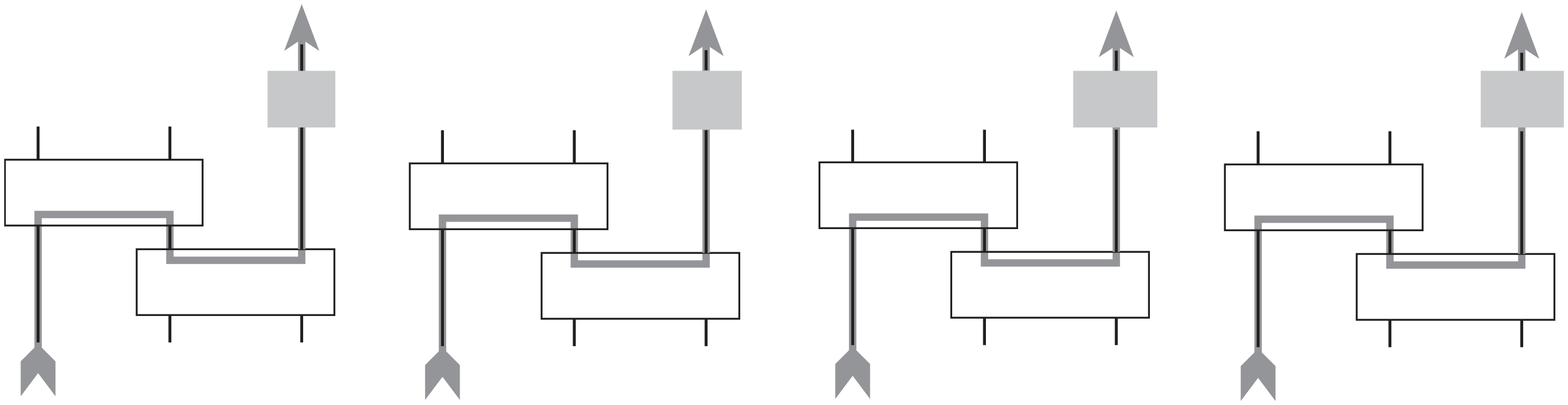,width=300pt}}{\footnotesize  

\begin{picture}(300,0)   
\put(41.6,31.5){${{\rm 1}}$}    
\put(117.1,31.5){${{\rm 1}}$}  
\put(196.6,31.5){${{\rm 1}}$}  
\put(273.1,31.5){${{\rm 1}}$}  
\put(17.6,50.6){${{\rm 1}}$}  
\put(92.5,50.5){${\beta_2}$}  
\put(170.7,50.8){${\beta_3}$}   
\put(248.5,50.5){${\beta_4}$}  
\put(54.3,66.9){${{\rm 1}}$}  
\put(130.1,67.8){${\gamma_2}$}  
\put(208.2,67.7){${\gamma_3}$} 
\put(284.9,67.8){${\gamma_4}$}     
\put(54.0,91.8){${{\phi_{out}=\phi_{in}}}$}  
\put(130.3,91.4){${\phi_{out}=\phi_{in}}$}  
\put(208.4,91.4){${\phi_{out}=\phi_{in}}$} 
\put(284.1,91.4){${\phi_{out}=\phi_{in}}$}     
\put(3.0,06.6){${\phi_{in}}$}  
\put(80.8,06.2){${\phi_{in}}$}  
\put(157.9,06.2){${\phi_{in}}$} 
\put(234.6,06.2){${\phi_{in}}$}     
\end{picture}}
\end{minipage}  

\vspace{-0mm}\noindent
where $\gamma_i\circ\beta_i$ has to be the identity so
$\gamma_i=\beta_i^{-1}$. These four pictures together yield the full
teleportation protocol!  The classical communication is encoded in
the fact that in each picture the unitary correction $\gamma_i$
depends on $\beta_i$, that is, the measurement outcome. Hence the classical communication does
not contribute to the transmission of the data, it only \em distributes the knowledge \em
about `which of the four pictures is actually taking place'.

To conclude this paragraph we stress that the functional labels are not actual
physical  operations but only arise in the above discussed mathematical isomorphism.
Further, in the generic example 

\vspace{5.5mm}\noindent{\footnotesize
\begin{minipage}[b]{1\linewidth}  
\centering{\epsfig{figure=eTeleport5.eps,width=60pt}}   

\begin{picture}(160,0)    
\put(90,27){$\!f_2$} 
\put(67.5,44.6){$\!\!f_1$}  
\put(52,4){\small$\phi_{in}$} 
\put(100,67){\small$\phi_{out}=(f_2\circ f_1)(\phi_{in})$}   
\end{picture}   
\end{minipage}}

\vspace{-0mm}\noindent
the order of the physical operations is opposite to the order in which their labels apply
to the input state in the expression $(f_2\circ f_1)(\phi_{in})$.
Algebraically,\footnote{The pictures really look much better than the formulas, don't
they?}
\[
k\cdot\zeta(\phi_{in}\otimes\Phi_{in})=\Psi_{f_1}\!\otimes(f_{\underline{\bf 2}}\circ
f_{\underline{\bf 1}})(\phi_{in})\ \ \ \ {\rm for}\ \ \ \
\zeta=(\PP_{f_{\underline{\scriptstyle\bf 1}}}\!\otimes
1)\circ(1\otimes\PP_{f_{\underline{\scriptstyle\bf 2}}})
\]
with $\Psi_f\stackrel{\simeq}{\longleftrightarrow} f$ and $\PP_f(\Psi_f)=\Psi_f$ as a new
notation. Slightly simpler, 
\[
(\PP_{f_{\underline{\scriptstyle\bf
1}}}\!\otimes 1)(\phi_{in}\otimes\Psi_{f_{\underline{\scriptstyle\bf
2}}})=\Psi_{f_1}\!\otimes(f_{\underline{\bf 2}}\circ
f_{\underline{\bf 1}})(\phi_{in})\,,
\]
by conceiving the first projector as a state. Furthermore, the above discussed $*$ in
${\cal H}_1^*\otimes{\cal H}_2$ which is  necessary to have a base-independent correspondence
with linear functions \em `is not a bug but a feature'\em, it actually witnesses (by means of a
phase conjugation) the fact
that the line changes its temporal direction every time it passes a projector box (see
\cite{Coe1}). 
Using the same line of thought it is also easy to reconstruct other
protocols such as \em logic-gate teleportation \em \cite{Gottesman} and
\em entanglement swapping \em \cite{Swap}, and, the quantum
information-flow interpretation also extends to multipartite
projectors.  We refer  the reader to \cite{Coe1,Coe2} for details on
this. Then we asked: 
\begin{center}{\bf ``Are these information-flow features specifically
related to the Hilbert space structure? Or  to ...''}
\end{center}

\paragraph{Sets, relations and the cartesian
product.}  Doesn't sound very `quantum' you say? 
Let's see. We make the following  substitutions in
the above:
\beqa
&{\rm Hilbert\ space\ } {\cal \ H}&\ \leadsto\ \ \ {\bf set\ } X\\
&{\rm linear\ function\ } f&\ \leadsto\ \ \ {\bf relation\ } R\\
&{\rm tensor\ product\ } \otimes&\ \leadsto\ \ \ {\bf cartesian\ product\ } \times
\eeqa
Can we also translate projectors to this world of relations? Observe that for projectors on
one-dimensional subspaces, which take the general form
$\PP_\psi=\langle \psi\mid-\rangle\cdot\!\!\mid\psi\rangle:{\cal H}\to{\cal H}$,
we have  
$\mid \psi\,\rangle\otimes\!\!\mid \psi\,\rangle
\  \stackrel{\simeq}{\longleftrightarrow}\
\langle\, \psi\mid\!-\rangle\cdot\!\!\mid \psi\,\rangle$,\footnote{Again we ignore
un-naturality, that is, the slight base-dependency.}   that is, projectors correspond with
symmetric pure tensors. By analogy we define a projector of type $X\!\to\! X$ as $A\times
A\subseteq X\times X$ in the world of relations.\footnote{Recall that a relation of type
$X\to Y$ is a subset of $X\times Y$ (cf.~its `graph').} Hence $R\times R\subseteq(X\times
Y)\times(X\times Y)$ with $R\subseteq(X\times Y)$ is a bipartite projector in the world of
relations which we denote by $\PP_R$ in analogy with $\PP_f$. Since for the identity
relation $1\subseteq X\times X$ we have 
$x_11x_2\Leftrightarrow x_1=x_2$ and since 
\[
\PP_R:=R\times R=\left\{\!\bigl((x_1,y_1),(x_2,y_2)\bigr)\in (X\times
Y)\times(X\times Y)\!\Bigm| x_1 R y_1,x_2 R y_2\right\}\!,
\]
for $R_1\subseteq X\times Y$ and $R_2\subseteq Y\times Z$ we
have\vspace{-1mm}
\[
(x_1,y_1,z_1)(1_X\!\otimes\PP_{R_2})(x_2,y_2,z_2)\ \Leftrightarrow\ y_1 R_2 z_1\ {\rm and}\
\underline{y_2 R_2 z_2},\ {\rm and},\ \underline{x_1=x_2}\,,
\]
\[
(x_2,y_2,z_2)(\PP_{R_1}\!\otimes 1_Z)(x_3,y_3,z_3)\ \Leftrightarrow\ \underline{x_2 R_1
y_2}\ {\rm and}\ x_3 R_1 y_3,\ {\rm and},\ \underline{z_2=z_3}\,.
\]
Setting $s_{in}:=x_1$, $s_{out}:=z_3$ and using the underlined 
expressions,
\[
(s_{in},y_1,z_1)\bigl((\PP_{R_{\underline{\scriptstyle\bf 1}}}\!\otimes
1)\circ(1\otimes\PP_{R_{\underline{\scriptstyle\bf 2}}})\bigr)(x_3,y_3,s_{out})
\]
entails $s_{in}(R_{\underline{\bf 2}}\circ R_{\underline{\bf
1}})s_{out}$. (we invite the reader to make a picture of this)~~{\bf But
this is not an accident!}

\section{The abstract algebra of entanglement} 

\paragraph{Categories for physical systems.} Which abstract structure do Hilbert
spaces and relations share? First of all, the above construction would not 
work if instead of relations we had taken functions. The importance of considering
appropriate maps indicates that we will have to consider \em categories\em.  As
theoretical computer scientists know, categories are not just a language, nor
metamathematics, nor hyper abstraction. They are mathematical objects in their own right
which arise very naturally in `real situations'. E.g.~one takes the state spaces of the
systems under consideration to be the \em objects\em, and (physical) operations on these
systems to be \em morphisms \em (including a \em skip \em operation), the axioms of a
category are then satisfied by the mere fact that operations can be \em composed\em.  We
denote by {\bf Rel} the category of sets and relations, by {\bf Set} the category of sets
and functions, by {\bf FdHilb} finite dimensional (complex) Hilbert spaces and linear
maps, and more generally, by ${\bf FdVec}_\mathbb{K}$ finite dimensional vector spaces
over a field
$\mathbb{K}$.

If instead of the cartesian
product we would have considered disjoint union on sets, again things wouldn't have worked
out.  Also in the quantum case the use of the tensor product is crucial. All this
indicates that we want some specific \em bifunctor \em $\boxtimes$ to live on our category,
$\times$ on {\bf Rel} and $\otimes$ on ${\bf FdVec}_\mathbb{K}$. Intuitively, we think of a
bifunctor as an operation which allows to combine systems, and also the operations thereon, and,
the \em bifunctoriality \em property has a clear physical interpretation: if $S_1$ and $S_2$ are
distinct physical entities, when performing operation
$O_1$ on $S_1$ and $O_2$ on $S_2$, the order in which we perform $O_1$ and $O_2$ doesn't matter.
One typically thinks of \em local operations \em on spatially separated systems.

In categories, \em elements \em of an object $A$ can be thought of as morphisms 
$q:\II\to A$ where
$\II$ is a unit for the bifunctor, i.e. $A\boxtimes\II\simeq\II\boxtimes A\simeq A$. In $({\bf
FdHilb},\otimes)$ we have
$\II:=\mathbb{C}$, and indeed,  maps $q: \mathbb{C}\to {\cal H}$ are in bijective correspondence
with ${\cal H}$ itself, by considering $q(1)\in{\cal H}$. In $({\bf
Set},\times)$ and $({\bf
Rel},\times)$ we have $\II:=\{*\}$, i.e., a singleton. In $({\bf
Set},\times)$ maps $q:\{*\}\to X$ are in bijective correspondence with elements of $X$
by considering $q(*) \in X$. But not in $({\bf Rel},\times)$! Morphisms $q\subseteq\{*\}\times
X$ now correspond to all subsets of $X$, which can be thought of as \em superpositions \em of
the individual elements.\footnote{Compare this to `superposition' in
lattice theoretic terms: an atomic lattice has superposition states if the join of two atoms has
additional atoms below it (e.g.~cf.~\cite{Coe0}).} 

We want not only a unit $\II$ for $\boxtimes$, but a full \em symmetric monoidal \em
structure, that is, we want the following \em natural isomorphisms\em\,\footnote{A categorical
isomorphism is a morphism
$f:A\to B$ with an inverse $f^{-1}:B\to A$, that is, $f\circ f^{-1}\!=1_A$ and
$f^{-1}\!\circ f=1_B$. A natural isomorphism is a strong notion of categorical isomorphism.
For vector spaces it essentially boils down to `base independent', e.g.~there exists a
natural isomorphism of type $({\cal H}_1^*\otimes {\cal H}_2)\longrightarrow ({\cal
H}_1\!\to\! {\cal H}_2)$  but not one of type $({\cal H}_1\otimes {\cal
H}_2)\longrightarrow ({\cal H}_1\!\to\! {\cal H}_2)$,
where we treat
${\cal H}_1\!\to\!{\cal H}_2$ as a Hilbert space.} 
\[
\lambda_A : A \simeq {\rm I}\boxtimes A\quad\quad\ \ \rho_A: A \simeq
A\boxtimes{\rm I}\quad\quad\ \ 
\sigma_{A,B}:A\boxtimes B\simeq B\boxtimes A 
\] 
\[
\alpha_{A,B,C}:A\boxtimes(B\boxtimes C)\simeq (A\boxtimes B)
\boxtimes C\,. 
\] 
Note here that we do not require $\boxtimes$-\em projections \em $p_{A,B}:A\boxtimes B\to A$ nor
$\boxtimes$-\em diagonals \em $\Delta_A:A\to A\boxtimes A$ to exist. More precisely, we don't
want them to exist, and this will be guaranteed by a piece of structure we shall
introduce. In physical terms this non-existence means  \em no-cloning \em \cite{WZ} and
\em no-deleting \em
\cite{Pati}.  In categorical terms it means that $\boxtimes$ is \em not a categorical
product\em.\footnote{See below where we discuss biproducs.} In logical terms this means
that we are doing
\em linear logic
\em
\cite{Girard,Lambek1,Lambek2,Seely} as opposed to classical logic. In linear logic we are not
allowed to copy and delete assumptions, that is, $A\wedge B\Rightarrow A$ and
$A\Rightarrow A\wedge A$ are not valid.

\vspace{-1.5mm}
\paragraph{Compact closure and information-flow.} Crucial in the analysis of the quantum
information-flow was ${\cal H}_1^*\otimes{\cal H}_2\simeq{\cal H}_1\!\to\!{\cal H}_2$.  In
categorical terms, making sense of ${\cal H}_1\!\to\!{\cal H}_2$ requires the category to be
\em closed\em.\footnote{For a monoidal category to be closed indeed means that we can
`internalize' morphism sets  $A\to B$ as objects, also referred to as the category having
exponentials. Typically, one thinks of $\boxtimes$ as conjunction and of this internalization as
implication.} To give sense to the
$*$ we require it to be
\em
$*$-autonomous
\em
\cite{Barr},\footnote{$*$-autonomy means that there exists an operation $*$ on the
monoidal category from which the internalization of morphism sets follows as $(A\boxtimes
B^*)^*$, cf.~classical logic where we have $A\Rightarrow B=\neg A\vee B= \neg(A\wedge \neg B)$
by the De Morgan rule.} and finally, requiring
${\cal H}_1^*\otimes{\cal H}_2\simeq{\cal H}_1\!\to\!{\cal H}_2$ implies that the category
is \em compact closed \em
\cite{Kelly}. In logical terms this means that we have the multiplicative fragment of linear
logic, with negation, and where \em conjunction is self-dual\em, that is, it coincides with
disjunction --- indeed, you read this correct, $A\wedge B\simeq A\vee B$. 

But we will follow a different path which enables us to use less categorical jargon.
This path is known in category theory circles as \em Australian \em or \em Max Kelly{\,} \em
style category theory. Although this style is usually conceived (even by category
theoreticians) as of an abstract$^\infty$ nature, in our particular case, it's bull's-eye
for understanding the quantum information-flow.\footnote{When we spell out this alternative
definition of compact closure it indeed avoids much of the categorical jargon. But it also
has a very elegant abstract formulation in terms of \em bicategories\em: a compact closed
category is a symmetric
monoidal category in which, when viewed as a one-object bicategory,
every one-cell $A$ has a left adjoint $A^*$.}

In \cite{KellyLaplaza} a category {\bf C} is defined to be compact closed iff  
for each object $A$ three additional pieces of data are specified, an object denoted \em
\em  $A^*$, a morphism $\eta_A:{\rm I}\to A^*\boxtimes A$ called \em unit \em 
and a morphism $\epsilon_A:A\boxtimes A^*\to {\rm I}$ called \em  
counit\em, which are such that the diagram
\begin{diagram}  
A&\rTo_{\simeq\ }&A\boxtimes{\rm
I}&\rTo_{1_A\boxtimes\eta_A}&A\boxtimes(A^*\boxtimes
A)\\ 
\dTo^{1_A}&&&&\dTo^{\simeq}\\ 
A&\lTo^{\simeq\ }&{\rm I}\boxtimes A&\lTo^{\epsilon_A\boxtimes
1_A}&(A\boxtimes A^*)\boxtimes A
\end{diagram}
and the same diagram for $A^*$ both commute. Although at
first sight this diagram seems quite intangible, we shall see that this
diagram perfectly matches the teleportation protocol. Both $({\bf
Rel},\times)$ and $({\bf FdVec}_\mathbb{K},\otimes)$ are compact closed,
respectively for $X^*:=X$,
$\eta_X=\{(*,(x,x))\mid x\in X\}$ and $\epsilon_X=\{((x, x),*)\mid x\in X\}$,
and, for $V^*$ the dual vector space of linear functionals, for
$\{\bar{e}_i\}_{i=1}^{i=n}$ being the base of $V^*$ satisfying
$\bar{e}_i(e_j)=\delta_{ij}$,
\[
\eta_V::1\mapsto\sum_{i=1}^{\!i=n\!}
\bar{e}_i\otimes e_i
\ \ \ \ \ \ \ \ {\rm and}\ \ \ \ \ \ \ \ \ 
\epsilon_V
::e_i\otimes\bar{e}_j\mapsto \delta_{ij}\,.\ \  
\]
(if
$V$ has an inner-product, $\bar{e}_i:=\langle e_i\mid-\rangle$)~~Note that $\eta_V(1)$ can
be thought of as an abstract generalization of the notion of an EPR-state.

Given the \em name \em and \em coname \em of a morphism $f:A\to B$, 
respectively
\[
\uu f\uuu:=(1\boxtimes f)\circ\eta_A:\II\to A^*\boxtimes B
\ \ \, {\rm and}\ \ \,
\dd f\ddd:=\epsilon_A\circ (f\boxtimes 1):A\boxtimes B^*\to\II,
\]
one can prove the
{\bf Compositionality Lemma} (\cite{AbrCoe2} \S3.3), diagrammatically,
\begin{diagram}  
A&\rTo_{\simeq\ }&A\boxtimes{\rm
I}&\rTo_{1_A\boxtimes\uu f_{\underline{\bf 2}}\!\uuu}&A\boxtimes(B^*\boxtimes
C)\\ 
\dTo^{f_{\underline{\bf 2}}\circ f_{\underline{\bf 1}}}&&&&\dTo^{\simeq}\\ 
C&\lTo^{\simeq\ }&{\rm I}\boxtimes C&\lTo^{\dd f_{\underline{\bf
1}}\!\ddd\boxtimes 1_A}&(A\boxtimes B^*)\boxtimes C
\end{diagram}
for $f_1:A\to B$ and $f_2:B\to C$. This lemma generalizes the defining
diagram of compact closedness since $\eta_A=\uu 1_A\!\!\uuu$ and
$\epsilon_A=\dd 1_{\!A}\ddd\ $ (cf.~EPR-state
$\stackrel{\simeq}{\longleftrightarrow} 1$).
The careful reader will have understood the picture by now,

\vspace{5.2mm}\noindent{ 
\begin{minipage}[b]{1\linewidth}  
\centering{\epsfig{figure=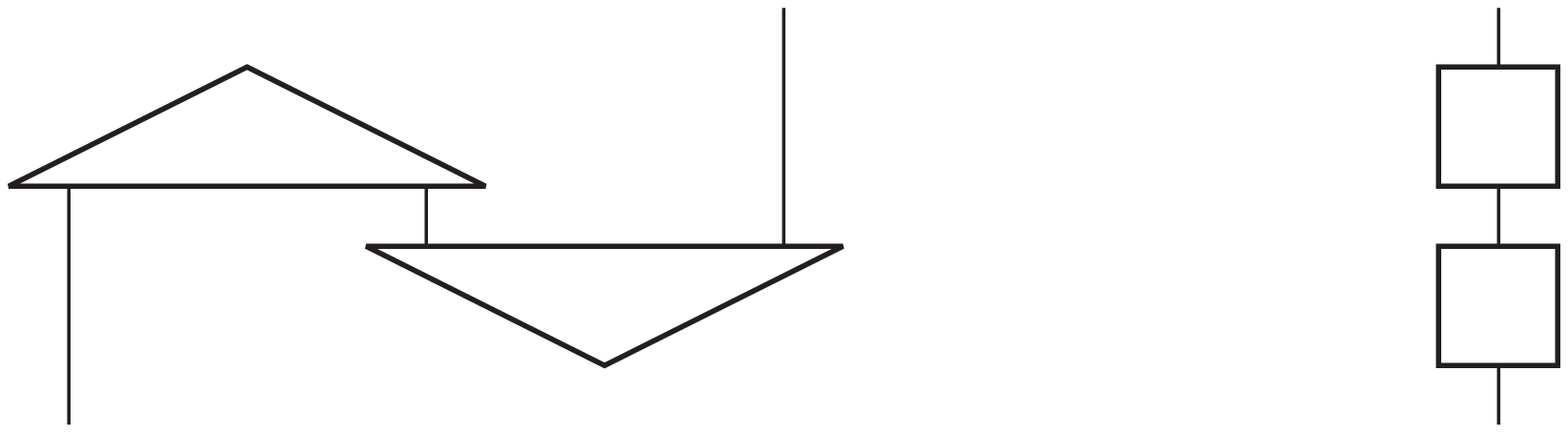,width=173.333pt}}  

\begin{picture}(173.333,0)        
\put(19.3,41.7){$\dd f_1\!\ddd$}
\put(59.1,24.0){$\uu f_2\!\uuu$} 
\put(122.3,30){\LARGE$=$}
\put(113.3,40){lemma}
\put(162,43){$f_2$} 
\put(162,23){$f_1$}
\put(26.3,54.7){$\II$}
\put(66.1,9.3){$\II$} 
\put(83.3,59.7){$C$}
\put(163.3,59.7){$C$}
\put(3.3,3.8){$A$}
\put(161.9,3.8){$A$}

\end{picture}
\end{minipage}}

\vspace{-0.5mm}\noindent
hence it seems as if there is an information flow through names and conames,

\vspace{2.2mm}\noindent{ 
\begin{minipage}[b]{1\linewidth}  
\centering{\epsfig{figure=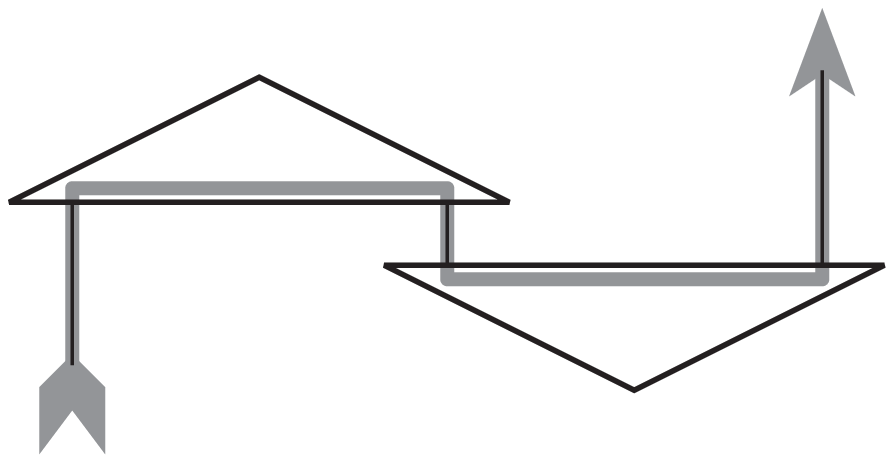,width=93.333pt}}    

\begin{picture}(93.333,0)        
\put(19.3,42.9){\small$\dd f_1\!\ddd$}
\put(58.1,22.9){\small$\uu f_2\!\uuu$} 
\end{picture} 
\end{minipage}}

\vspace{-4.7mm}\noindent
Are we really there yet? 
We actually have two things, names and conames, and
names act as `the output of a bipartite projector'
while conames act as `the input of a bipartite
projector'. The obvious thing to do is  to glue a
coname and a name together in order to produce a
bipartite projector.

\vspace{2.5mm}\noindent{ 
\begin{minipage}[b]{1\linewidth}  
\centering{\epsfig{figure=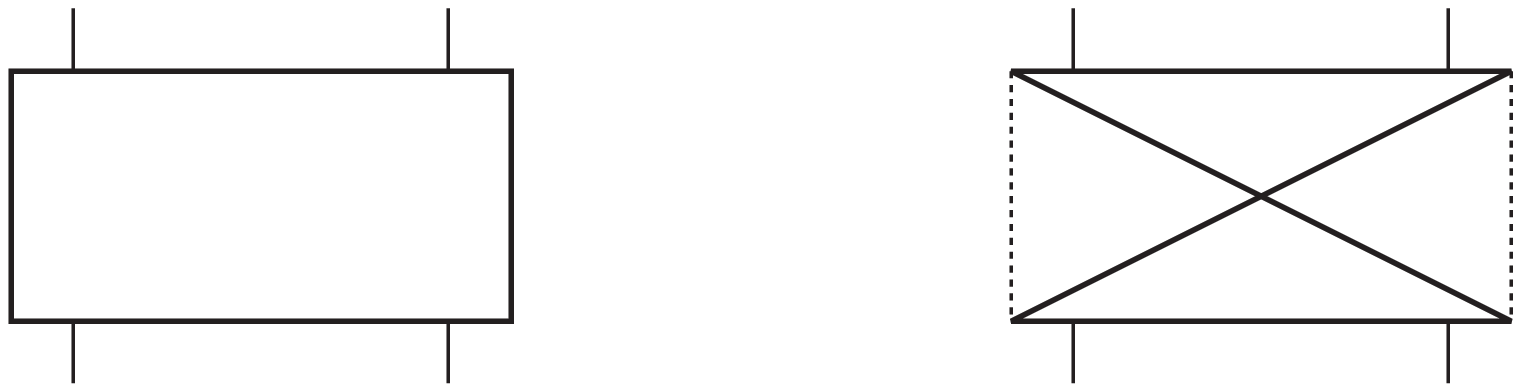,width=150pt}}  

\begin{picture}(150,0)        
\put(22,28){\large$\PP_{\!f}$} 
\put(68,28){\LARGE$:=$}
\put(116.5,35.2){$\uu f\uuu$}
\put(116.5,21.5){$\dd f\ddd$}
\end{picture}
\end{minipage}}

\vspace{-3.5mm}\noindent 
However, we have a type-mismatch.
\[
\PP_f:\stackrel{?}{=}\uu f\uuu\circ\dd f\ddd:A\boxtimes 
B^*\to A^*\!\boxtimes B
\]
To solve this problem we
need a tiny bit of extra structure. This bit of extra structure will
capture the idea of \em complex conjugation\em. When conceiving
elements as \em Dirac-kets\em, it will provide us with a notion of \em Dirac-bra\em.
We will introduce strong compact closure, metaphorically,
\[
{{\sf strong\ compact\ closure}\over{\sf compact\ closure}}
\ \simeq\
{{\sf sesquilinear\ inner\mbox{\sf-}product\ space}\over{\sf vector\ space}}\,.
\]

\paragraph{Strong compact closure, inner-products and projectors.} The assignment
$A\mapsto A^*$ which arises as part  of the definition of compact closure actually
extends to one on morphisms,
\begin{diagram}  
B^*\!&\rTo_{\simeq\ \ }&{\rm I}\boxtimes B^*\!&\rTo_{\ \eta_A\boxtimes
1_{B^*}}&(A^*\boxtimes A)\boxtimes B^*\\ 
\dTo^{f^*}&&&&\dTo_{1_{A^*}\!\boxtimes f\boxtimes 1_{B^*}\hspace{-1.3cm}}\\   
A^*\!&\lTo^{\simeq\ }&A^*\!\boxtimes {\rm I}&\lTo^{\ 1_{A^*}\!\boxtimes
\epsilon_B}&A^*\boxtimes (B\boxtimes B^*)
\end{diagram}
and again this looks much nicer in a picture,

\vspace{4.5mm}\noindent{ 
\begin{minipage}[b]{1\linewidth}  
\centering{\epsfig{figure=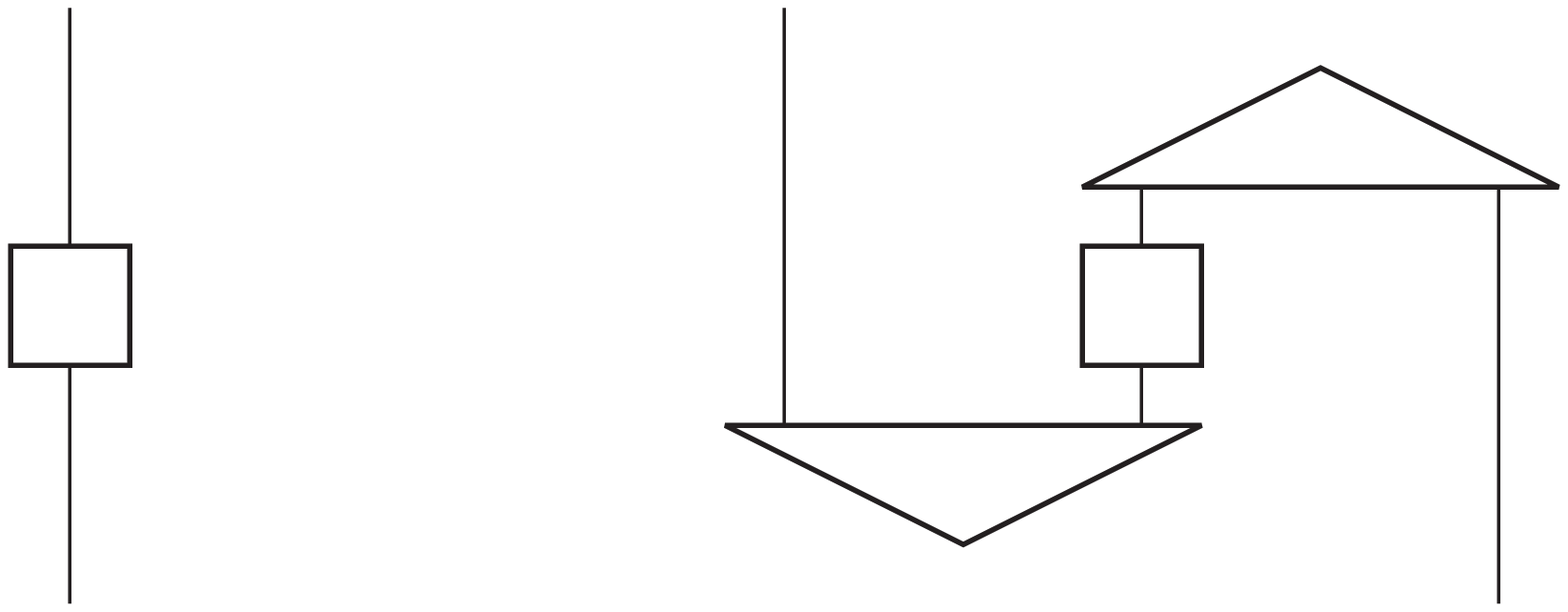,width=173.333pt}}  

\begin{picture}(173.333,0)        
\put(98.8,23.6){$\uu 1\uuu$}
\put(138.6,61.1){$\dd 1\ddd$} 
\put(42.3,42){\LARGE$=$}
\put(38.2,42.4){\LARGE$:$}
\put(39.7,52){def.}
\put(2.5,43){$f^*$} 
\put(122,43){$f$} 
\put(144.1,74.7){$\II$}
\put(104.3,9.3){$\II$} 
\put(3.3,79.7){$A^*$}
\put(82.3,79.7){$A^*$}
\put(3.3,3.8){$B^*$}
\put(160.9,3.8){$B^*$}

\end{picture}
\end{minipage}}

\vspace{-1.5mm}\noindent
It is as if the information flows backward through $f$,

\vspace{1.5mm}\noindent{ 
\begin{minipage}[b]{1\linewidth}  
\centering{\epsfig{figure=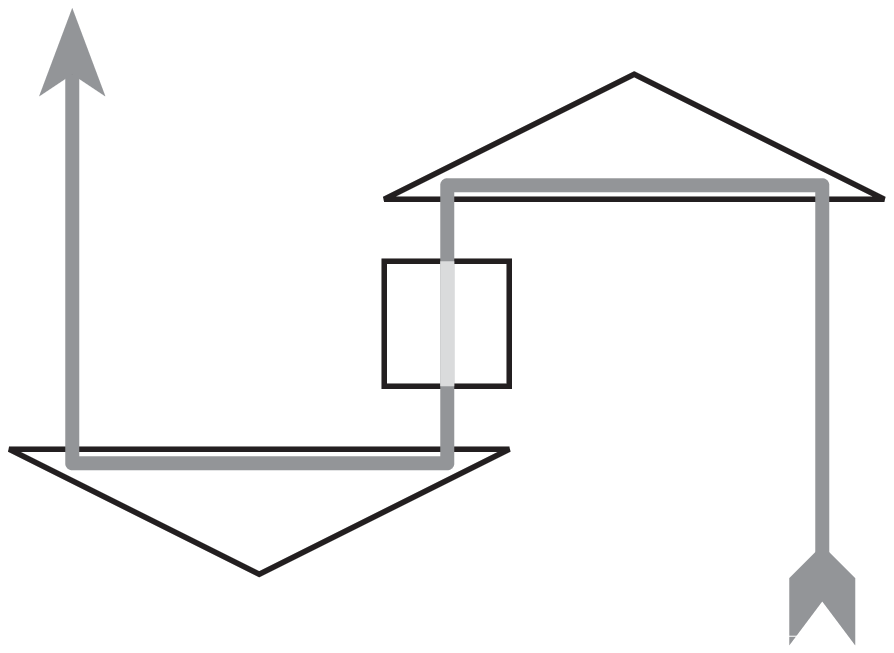,width=93.333pt}}  

\begin{picture}(93.333,0)        
\put(20.0,23.5){$\uu 1\uuu$}
\put(58.7,61.3){$\dd 1\ddd$} 
\put(44,43){$f$} 
\end{picture}
\end{minipage}}

\vspace{-4.5mm}\noindent
For vector spaces the matrix of $f^*$ is the \em transposed
\em of the matrix of $f$ when taking
$\{\bar{e}_i\}_{i=1}^{i=n}$ as base for $V^*$ given base
$\{{e}_i\}_{i=1}^{i=n}$  of $V$\,. For relations $R^*$ is
the \em relational converse \em of
$R$.  One verifies that
$(\ )^*:{\bf C}\to {\bf C}$ is a
\em contravariant functor\em , that is $(f_1\circ
f_2)^*=f_2^*\circ f_1^*$, and that there exists a natural
isomorphisms $A^{**}\!\simeq A$ and $(A\boxtimes B)^*\simeq A^*\boxtimes
B^*$.
\begin{definition}\em
A \em strongly compact closed category \em
\cite{AbrCoe2} is a compact closed category for which $A=A^{**}$ and for
which the assignment
$A\mapsto A^*$ and $(A\boxtimes B)^*=A^*\boxtimes B^*$ has also an
\em involutive covariant functorial extension\em, which  commutes 
with the compact closed structure.  
\end{definition}

We set $f\mapsto f_*$ for this functorial extension.
For each morphism $f:A\to B$ we 
define its \em adjoint \em and a \em bipartite projector \em as
\[
f^\dagger:=(f_*)^*=(f^*)_*:B\to A
\]
and now we can define bipartite projectors to be
\[
\PP_f:=\uu f\uuu\circ(\uu f\uuu)^\dagger=\uu f\uuu\circ\dd
f_*\ddd:A^*\!\boxtimes B\to A^*\!\boxtimes B,
\]
and we call an isomorphism $U:A\to B$ \em unitary \em iff $U^{-1}=U^\dagger$.  An abstract
notion of \em inner-product \em also emerges. Given
elements
$\psi,\phi:\II\to A$  we set
$\langle\psi\mid\phi\rangle:=\psi^\dagger\circ\phi\,\in\CC(\II,\II)$ where $\CC(\II,\II)$ are
the morphisms of type $\II\to \II$ --- we discuss these \em scalars \em in more detail
below. We can now prove the usual defining properties of adjoints and unitarity in abstract
generality,
\[
\langle
f^\dagger\!\circ\psi\mid\phi\rangle_B=(f^\dagger\!\circ\psi)^\dagger\!\circ\phi=
\psi^\dagger\!\circ f\circ\phi=\langle \psi\mid 
f\circ\phi\rangle_A\,,
\]
\[
\langle U\circ\psi\mid U\circ\varphi\rangle_B=
\langle U^\dagger\!\circ U\circ\psi\mid \varphi\rangle_A=
\langle \psi\mid \varphi\rangle_A\,.
\]
When calling
$\psi:\II\to A$ a \em ket\em, then $\psi^\dagger:\II\to A$ is the corresponding \em bra \em
and the scalar
$\phi^\dagger\circ \psi:\II\to\II$ is a \em bra-ket\em. Hence strong compact closure provides a
nice and juicy lump of Hilbert space --- see \cite{AbrCoe2} \S7 and
\cite{AbrCoe3} \S2 for details. 

The category $({\bf Rel},\times)$ is trivially strongly compact closed for $R_*:=R$, so
$R^\dagger=R^*$, that is, adjoints are relational converses. The same goes for any compact
closed category where $A^*=A$. For $({\bf FdVec}_\mathbb{K},\otimes)$ we don't have $V^*= V$,
nor does the above defined compact closed structure satisfy $V^{**}=V$, so it
cannot be extended to a strong compact closed structure.  But for
$\mathbb{K}:=\mathbb{R}$, finite-dimensional real inner-product spaces are strongly compact
closed for
$V:=V^*$ and
$\epsilon_V:=\langle-\!\mid\!-\rangle$, and for $\mathbb{K}:=\mathbb{C}$, our main category
$({\bf FdHilb},\otimes)$ is also strongly compact closed when we take
${\cal H}^*$ to be the \em conjugate space\em, that is, the Hilbert space with the same
elements as
${\cal H}$ but with $\alpha \sdot_{\HH^*} \phi := \bar{\alpha} \sdot_{\HH} \phi$ as scalar
multiplication and $\langle \phi \mid \psi \rangle_{\HH^*} := \langle \psi \mid \phi
\rangle_{\HH}$ as (sesquilinear) inner-product.
We can then set 
$\epsilon_\HH:{\cal H}\otimes{\cal
H}^*\to\II::\psi\otimes\phi\mapsto\langle\phi\mid\psi\rangle$. One verifies that we recover the
usual notion of adjoint, that is, the \em conjugate transpose\em, where $(\ )^*$ provides
transposition while $(\ )_*$ provides complex conjugation.

Let us end this paragraph by saying that most things discussed above extend to \em infinite
dimensional \em settings when using ideas from
\cite{ABP}.

\paragraph{A note on categorical traces.} This paragraph slightly
diverges from our  story line, but we do want to mention that much of
the inspiration for \cite{Coe1,Coe2}  emerged from \cite{AbrCoe1} where
we studied the physical realization of `abstract traces' \cite{JSV},
which generalize traditional \em feedback traces \em
\cite{Abr,Bai}. It turns out that both on $({\bf Rel},\times)$ and $({\bf
FdVec}_\mathbb{K},\otimes)$, due to compact closure,  the trace also admits a feedback-loop type
interpretation, but a linear `only-use-once' one. Please consult \cite{AbrCoe3} for more details
and some nice pictures.

\section{Beyond von Neumann's axiomatics}

\paragraph{Biproducts.} Strong compact closure provides a serious lump of Hilbert space, but we
need some additional types which enable to encode classical information
and its flow in our quantum formalism. They will capture `gluing pictures
together' and `distributing the knowledge on in which picture we are'
(cf.~\S2). To this means we use \em biproducts\em, that is,   objects
$A\boxplus B$ which both are the \em product \em and the \em coproduct
\em for $A$ and $B$, and corresponding induced morphisms $f\boxplus g:A\boxplus B\to C\boxplus D$ for
$f:A\to C$ and $g:B\to D$.
Contrary to $\boxtimes$, biproducts go (by
definition) equipped with
\em projections \em $p_j:\boxplus_i  A_i\to A_j$, also with \em injections  
\em $q_j: A_j\to\boxplus_i  A_i$,
and with \em pairing \em and \em copairing \em operations, 
$\langle f_i\rangle_i:A\to \boxplus_i  A_i$ and $[f_i]_i:\boxplus_i  A_i\to A$,
for morphisms $f_i:A\to A_i$ and $g_i:A_i\to A$ with coinciding domain and codomain respectively. From
these we can construct \em diagonals \em and \em codiagonals\em,
$\Delta_A:=\langle 1_A,1_A\rangle:A\to A\boxplus A$ and $\nabla_A:=[ 1_A,1_A]:A\boxplus A\to
A$.  This `non-linear' $\boxplus$-structure encodes that there is no difference between looking at two
pictures separately, or together --- the components of a compound
quantum system cannot be considered separately, hence $\boxtimes$ is linear.

We take the projections and injections such that they work nicely together with the strong compact
closure by setting $q_i^\dagger=p_i$ (and hence $p_i^\dagger=q_i$). 
Of crucial importance for us is the 
\em distributivity \em of $\boxtimes$ over $\boxplus$,\footnote{Which follows by closedness of 
$\boxtimes$ and $\boxplus$ being a coproduct.} that is, 
there is a natural isomorphism
\[
{\sf\scriptstyle DIST}:A\boxtimes(B_1\boxplus B_2)\simeq(A\boxtimes B_1)\boxplus(A\boxtimes B_2)\,.
\]
For $({\bf Rel},\times)$ the \em disjoint union \em $+$ provides a
biproduct structure with inclusion as injections. For $({\bf
FdHilb},\otimes)$ the \em direct sum \em $\oplus$ provides a biproduct
structure with coordinate projections as projections. 

\vspace{-1.5mm} 
\paragraph{Categorical quantum mechanics.}  We define a quantum formalism relative to any strongly
compact closed category with some biproducts.

{\bf i.}~We take \em state spaces \em to be objects which do not involve explicit
biproducts and use
$\boxtimes$ to describe compound systems. The basic \em data unit \em is a state space $Q$ which is
unitary isomorphic to $\II\boxplus\II$, which in the case of $({\bf Rel},\times,+)$ where 
$\II\boxplus\II=\{*\}\!+\!\{*\}$ yields the \em boolean type \em and in the case of $({\bf
FdHilb},\otimes,\oplus)$ where $\II\boxplus\II=\mathbb{C}\oplus\mathbb{C}$ yields the \em
qubit type\em.

{\bf ii.}~Explicit biproducts express `different pictures' due to distinct measurement outcomes, they
enable to encode \em classical data\em. The distributivity isomorphism
${\sf\scriptstyle DIST}$ expresses \em exchange of classical data\em\,!  (see below)

{\bf iii.}~We have already defined bipartite projectors. To turn them into a measurement we
need to glue a complete family of mutually orthogonal ones to each other. More generally,
we define a \em spectral decomposition \em to be a unitary morphism $U:A\to\boxplus_i 
A_i$. We define the corresponding \em non-destructive measurement \em  
to be the copairing 
\[
\langle\PP_i \rangle_{i}:A\to \boxplus_i  A
\ \ \ \ {\rm where}\ \ \ \
\PP_j=\pi_j^\dagger\circ\pi_j:A\to A
\ \ \ \ {\rm for}\ \ \ \
\pi_j=p_j\circ U
\]
with $p_j: \boxplus_i  A_i\to A_j$ the projections for the biproduct $\boxplus_i  A_i$. As
shown in
\cite{AbrCoe2}, these general \em projectors \em  $\PP_i:A\to A$ are self-adjoint, mutually orthogonal,
and their sum is $1_A$ --- we discuss the \em sum of morphisms \em below. When the spectral
decomposition is of type $A\to\boxplus_i  \II$ the corresponding measurement is \em
non-degenerated\em. We call such a spectral decomposition, which by the defining property
of products can be rewritten as $\langle\pi_i
\rangle_{i}:A\to\boxplus_i 
\II$, a \em non-degenerated destructive measurement\em.
For an explicit definition of an \em abstract Bell-base measurement\em, or any other measurement which
allows teleportation, we refer to \cite{AbrCoe2}. \em Isolated
reversible dynamics \em is unitary.

{\bf iv.}~The passage from a non-degenerated non-destructive measurement to a destructive one
involves dropping $\psi_i:=\pi_i^\dagger:\II\to A$. We conceive such a component as a
\em preparation\em. Hence a non-destructive measurement decomposes in $\langle\pi_i
\rangle_{i}$, which gives the measurement's outcome, and
$\psi_i$, which gives the state `after the collapse' (cf.~von 
Neumann's projection postulate).

\paragraph{Abstract quantum teleportation.} The righthandside of the
diagram 

\vspace{2.5mm}\noindent 
\begin{diagram} 
Q&\rIs&Q\\
&&\dTo^{(1\boxtimes \uu 1_Q\!\uuu)\circ\rho_Q}&\hspace{-1.5cm}{\bf produce\ 
EPR\mbox{\bf
-}pair}\\ &&Q\boxtimes(Q^*\!\boxtimes Q)\\
&&\dTo^{\simeq}&\hspace{-1.5cm}{\bf spatial\ relocation}\\
&&(Q\boxtimes Q^*)\boxtimes Q\\
\dTo^{\langle\, 1_Q\rangle_{i=1}^{i=4}}&&\dTo^{\quad\quad\bigl\langle \dd 
\beta_i\ddd\bigr\rangle_{i=1}^{i=4}\boxtimes 1_Q}&\hspace{-1.5cm}{\bf
Bell\mbox{\rm -}base\ measurement}\\ &&\left(\boxplus_{i=1}^{i=4}{\rm
I}\right)\!\boxtimes Q\\
&&\dTo^{\ \ \ \ \ \ \qquad(\boxplus_{i=1}^{i=4}\lambda^{-1}_Q)\!\circ{\sf\scriptstyle
DIST}}&\hspace{-1.5cm}{\bf
\quad classical\ communication\quad}\\ &&\ \ \ \boxplus_{i=1}^{i=4}Q\\
&&\dTo^{\boxplus_{i=1}^{i=4}\beta_i^{-1}}&\hspace{-1.5cm}{\bf
unitary\ correction}\\
\ \ \ \ \boxplus_{i=1}^{i=4} Q&\rIs&\boxplus_{i=1}^{i=4} Q\!\!\!\!\!\!
\end{diagram}

\vspace{1mm}\noindent 
\em gives a complete description of the teleportation protocol\em. The
lefthandside expresses the intended behavior (obtaining an identity in each of the four
pictures).  In \cite{AbrCoe2} we proved correctness, \em the diagram commutes\,\em!

Abstract presentations and proofs of correctness of \em logic gate
teleportation \em
\cite{Gottesman} and \em entanglement swapping \em
\cite{Swap} can be found in \cite{AbrCoe2}.

Immediately after the Bell-base measurement the type is $\left(\boxplus_{i=1}^{i=4}{\rm
I}\right)\!\boxtimes Q$ where $\boxplus_{i=1}^{i=4}{\rm
I}$ represents the four different measurement outcomes.  However, these four
pictures only exist `locally'. After \em distributing this information\em,
\begin{diagram}
\left(\boxplus_{i=1}^{i=4}{\rm
I}\right)\!\boxtimes
Q&\rTo^{\ \ \ {\sf\scriptstyle
DIST}\ \ \ }&\boxplus_{i=1}^{i=4}(\II\boxtimes
Q)&\rTo^{\boxplus_{i=1}^{i=4}\lambda^{-1}_Q}&\boxplus_{i=1}^{i=4}Q\,,
\end{diagram}
there are four different pictures `globally'. Hence
we can apply the appropriate unitary correction
$\beta_i^{-1}:Q\to Q$ in each picture, that is,
$\boxplus_{i=1}^{i=4}\beta_i^{-1}$.

The spectrum of a measurement $\langle\PP_i \rangle_{i}$ is the index set
$\{i\}_i$, which for example could encode locations in physical space.  Since for
teleportation we assume to work with spatially located particles, that is, there are no
spatial superpositions, the associativity natural isomorphism allows to encode
\em spatial association \em (i.e.~proximity) in a qualitative manner.

\vspace{-1.5mm}
\paragraph{Scalars, normalization, probabilities and the Born rule.} Up to now one
might think that the abstract setting is purely qualitative (whatever that means
anyway).  But it is not!  The scalars ${\bf C}(\II,\II)$ of any monoidal category
${\bf C}$ have a commutative composition \cite{KellyLaplaza}, that is,
a \em multiplication\em.  

If the biproduct $\II\oplus\II$ exists, 
we can define a \em sum of scalars \em $s,s':\II\to\II$ as 
\[
s+s':=\nabla_\II\circ(s\boxplus s')\circ\Delta_\II:\II\to\II\,
\]
and one shows that the above defined multiplication distributes
over this sum and that there is a zero
$O_\II:\II\to\II$.   Hence we obtain an \em abelian semiring\em.\footnote{That
is, a field except that there are no inverses for addition nor for
multiplication.}

Furthermore, each scalar $s:\II\to\II$ induces a natural transformation
\[
s_A:\lambda^{-1}_A\circ (s\otimes 1_A)\circ \lambda_A:A\to A
\]
for each object $A$, which allows us to define \em scalar
multiplication \em as $s\sdot f:=f\circ s_A$ for $f:A\to
B$, where $f\circ
s_A=s_B\circ f$ by naturality, that is, \em morphisms
preserve scalar multiplication\em.  

Since we have an inner-product (which, of course, is scalar valued) we can now talk
about \em normalization \em e.g.~an element $\psi:\II\to A$ is normalized
iff $\psi^\dagger\circ\psi=1_\II$.\footnote{A discussion of normalization of projectors
can be found in \cite{AbrCoe3}.} Besides the special scalars $1_\II$ and $0_\II$ there
are many others, those which satisfy $s^\dagger=s$, those of the form
$s^\dagger\circ s$, those which arise from inner-products of normalized elements, and
the latter multiplied with their adjoint, in $({\bf FdHilb},\otimes,\oplus)$
respectively being $1$ and $0$, the reals $\mathbb{R}$, the positive reals
$\mathbb{R}^+$, the unit disc in $\mathbb{C}$ and the unit interval $[0,1]$.

Consider now the basic protocol of (non-destructively) measuring a state  
\begin{diagram}
\II & \rTo^{\psi} & A & \rTo^{\langle \PP_i \rangle_{i=1}^{i=n}}
&\boxplus_{i=1}^{i=n}A\,.
\end{diagram}
If we look at one component of the biproduct, i.e., one picture,
\begin{diagram}
\II & \rTo^{\psi} & A & \rTo^{\pi_i}
&\II & \rTo^{\pi_i^\dagger}
&A\,,
\end{diagram}

\begin{picture}(80,0)   
\put(115,16){$\underbrace{\mbox{\hspace{4.5cm}}}$}
\put(177,0){$s_i\in{\bf
C}(\II,\II)$} 
\end{picture}

\vspace{2.5mm}\noindent
we discover a special scalar of the
`unit disc type'. One verifies that 
\[
{\sf\scriptstyle PROB}(\PP_i,\psi):=s_i^\dagger\circ
s_i
\ \ \ \ \ \ {\rm satisfies}\ \ \ \ \ \
\sum_{i=1}^{i=n}{\sf\scriptstyle
PROB}(\PP_i,\psi)=1_\II\,,
\]
hence these `$[0,1]$ type' scalars
${\sf\scriptstyle PROB}(\PP_i,\psi)$ provide an
abstract notion of \em probability \em
\cite{AbrCoe2}. Moreover, using our abstract
inner-product one verifies that 
${\sf\scriptstyle PROB}(\PP_i,\psi)=\langle \psi
\mid\PP_i\circ\psi\rangle$, that is, we prove the
\em Born rule\em.

\vspace{-1.5mm}
\paragraph{Mixing classical and quantum uncertainty.} This section comprises a \em
proposal \em for the abstract status of density matrices. Having only one page left,
we need to be brief. In the von Neumann formalism density matrices are required for two
reasons: {\bf i.}~to describe part of a larger (compound) system, say \em ontic density
matrices\em, and,  {\bf ii.}~to describe a system about which we have incomplete knowledge,
say \em epistemic density matrices\em.   Hence ontic density
matrices arise by considering one component of an element of the name type, $\uu
\xi\uuu:\II\to A_1\boxtimes A_2$ for $\xi:A_1^*\to A_2$.
In order to produce  epistemic density matrices,
consider the situation of a measurement, but we extract the information concerning the
actual outcome from it, that is, we do the converse of distributing classical data,
\begin{diagram}
\II  \rTo^{\!\!\!\!\!\phi\!\!\!\!\!} & A & \rTo^{\langle \PP_i \rangle_{i}\,}
&\boxplus_{i}A& \rTo^{\boxplus_{i}\lambda_Q}
&\boxplus_{i}(\II\boxtimes A)& \rTo^{\ {\sf\scriptstyle DIST}^{-1}}
&(\boxplus_{i}\II)\boxtimes A\,.
\end{diagram}
This results again in an element of the name type, $\uu
\omega\uuu:\II\to (\boxplus_{i}\II)\boxtimes A$ for
$\omega:(\boxplus_{i}\II)^*\to A$. Metaphorically one could say that the \em
classical data is entangled with the quantum data\em. Since our formalism allows both to
encode classical data and quantum data there is no need for a separate density matrix
formalism as it is the case for the von Neumann formalism.

One verifies that the principle of \em no signalling faster than light \em
still holds for the name type in the abstract formalism, that is, operations locally on
one component will not alter the other, provided there is no classical data
exchange. But there can be a passage from ontic to epistemic e.g.
\[
\II\to A_1\boxtimes A_2\ \ \ \leadsto\ \ \ \II\to (\boxplus_{i}\II)\boxtimes A_2 
\]
when performing the measurement $\langle \pi_i
\rangle_{i}\boxtimes 1_{A_2}:A_1\boxtimes A_2\to
(\boxplus_{i}\II)\boxtimes A_2$.  For epistemic density matrices this means that the
classical data and the  quantum data are truly distinct entities.

Using Lemma 7.6 of \cite{AbrCoe2} one verifies that 
$\omega_*:\boxplus_{i}\II\to A^*$ is given by $\omega_*=[s_i\sdot(\pi_i^*\circ
u_\II)]_{i}$ where
$s_i:=\pi_i\circ\phi$ and
$u_\II:\II\simeq\II^*$  (a natural transformation which exists by compact closure).
Hence $\omega$ and hence also $\uu\omega\uuu$ is determined by a list of orthogonal (pure)
states $(\pi^\dagger_i:\II\to A)_{i}$ and a list of scalars $(s^\dagger_i:\II\to
A)_{i}$ all of the unit disc type --- compare this to the orthogonal
eigenstates of a standard Hilbert space density matrix and the corresponding eigenvalues 
which all are of the $[0,1]$ type.   
 
So we can pass from pure states $\phi:\II\to \II$
to density matrices by `plugging in an ancilla', which either represents classical data
(epistemic) or which represents an external part of the system (ontic). The other
concepts that can be derived from basic quantum mechanics by `acting on part of a
bigger system' (\em{non-isolated dynamics}\em, \em generalized measurements\em, \cite{HK}
etc.) can also be defined abstractly, e.g.~\emph{generalized
measurements} as
\[
\langle f_i \rangle_{i=1}^{i=n}:A\to \boxplus_i  A
\qquad {\rm with}\qquad
\sum_{i=1}^{i=n} f^\dagger_i\circ f_i=1_A\,,
\]
while abstract
analogous of theorems such as Naimark's can be proven. Of course, many
things remain to be verified such as abstract analogous of Gleason's theorem.  
I might have something to add to this in my talk.
\texttt{:)}  

\end{document}